\documentclass[useAMS,usenatbib]{mn2e}
\usepackage{graphicx}
\usepackage[latin1]{inputenc}
\usepackage[portuges]{babel}
\usepackage{subfigure}
\usepackage{indentfirst}
\usepackage{placeins}
\usepackage{multicol}
\usepackage{Times} 
\usepackage{textcomp}

\title[Stable retrograde orbits around the triple system 2001 SN263]
{Stable retrograde orbits around the triple system 2001 SN263}
\author[Araujo, R.A.N.; Winter, O.C.; Prado, A.F.B.A]
{R.A.N.Araujo$^{1}$\thanks{E-mail:ran.araujo@gmail.com}, O.C.Winter$^{1,2}$\thanks{E-mail:ocwinter@gmail.com}, A.F.B.A.Prado$^{2}$\thanks{E-mail:prado@dem.inpe.br} \\
$^1$UNESP - Univ Estadual Paulista, Grupo de Din\^amica Orbital e Planetologia, CEP 12516-410, Guaratinguet\'a, SP, Brazil.\\
$^2$ INPE - National Institute for Space Research, CEP 12201970, Sao Jose dos Campos, SP, Brazil}
\begin{document}

\date{}

\pagerange{\pageref{firstpage}--\pageref{lastpage}} \pubyear{2015}
\maketitle
\label{firstpage}

\begin{abstract}
The NEA 2001 SN263 is a triple system of asteroids and it is the target of the ASTER MISSION - First Brazilian Deep Space Mission. 
The announcement of this mission has motivated a 
study aimed to characterize regions of stability of the system. \cite{b3}, characterized the stable regions around the components of the 
triple system for the planar and prograde cases. Through numerical integrations they found that the stable regions are in two tiny internal 
zones, one of them placed very close to Alpha and another close to Beta, and in the external region. For a space mission aimed to place the 
probe in the internal region of the system those results do not seem to be very interesting. Therefore, knowing that the retrograde orbits 
are expected to be more stable, here we present a complementary study. We now considered particles orbiting the components of the system, in 
the internal and external regions, with relative inclinations between $90^{\circ}< I \leqslant180^{\circ}$, i.e., particles with retrograde orbits. 
Our goal is to characterize the stable regions of the system for retrograde orbits, and then detach a preferred region to place the space probe. 
For a space mission, the most interesting regions would be those that are unstable for the prograde cases, but stable for the retrograde cases. 
Such configuration provide a stable region to place the mission probe with a relative retrograde orbit, and, at the same time, guarantees a 
region free of  debris since they are expected to have prograde orbits. 
We found that in fact the internal and external stable regions significantly increase when compared to the prograde case. 
For particles with $e=0$ and $I=180^{\circ}$, we found that nearly the whole region around Alpha and Beta remain stable. 
We then identified three internal regions and one external region that are very interesting to place the space probe. 
We present the stable regions found for the retrograde case and a discussion on those preferred regions. 
We also discuss the effects of resonances of the particles with Beta and Gamma, and the role of the Kozai mechanism in this scenario. 
These results help us understand and characterize the stability of the triple system 2001 SN263 when retrograde orbits are considered, and 
provide important parameters to the design of the ASTER mission. 

\end{abstract}


\section{Introduction}

\begin{table*}
\centering
\label{tab_elements}
\begin{minipage}{120mm}
\centering
\caption{Physical and orbital data of the three components of the system 2001 SN263}
\end{minipage}
\begin{tabular}{|c|c|c|c|c|c|c|}
\hline
Body		&Mass$^{(1)}$  				&Radius  		&Period$^{(1)}$    &a$^{(1)}$    &e$^{(1)}$	 &I$^{(1)}$ $^{(*)}$\\
\hline
Alpha		&$917.47\times10^{10}$kg		&$1.3$ km $^{(1)} $     &$2.80$ years      &		 &		  &		\\ 						       \\
\hline
Beta		&$24.04\times10^{10}$kg		 	&$0.39$ km$^{(2)}$     &$6.23$ days      &$16.63$ km     &$0.015$   	 & $0.0^{\circ}$         \\
\hline
Gamma		&$9.77\times10^{10}$ kg 		&$0.29$ km$^{(2)}$     &$ 0.69$ days	  &$3.80$ km     &$0.016$ 	 &$\approx14^{\circ}$   \\
\hline
\multicolumn{7}{l}{ 1 - \cite{b5}.}\\
\multicolumn{7}{l}{2 - \cite{b3}}\\
\multicolumn{7}{l}{(*) With respect to the equator of Alpha.}\\
\end{tabular}
\end{table*}

The population of NEAs (Near-Earth Asteroids) is composed by asteroids orbiting the Sun with perihelion $q\leqslant1.3$ AU.
They are classified according to their orbital characteristics as: Amor ($a>1$ AU and $1.017\leq q \leq 1.3$ AU), 
Apollo ($a>1$ AU and $q\leq1.017$ AU), Atens ($Q>0.983$ AU) and IEO - interior to the Earth's orbit ($Q<0.983$ AU), where $a$ is 
the semi-major axis and $Q$ is the aphelion distance of the orbit of the asteroid.

The population of NEAs is composed mostly by asteroids coming from the Main-Belt asteroids. 
According to \cite{b1}, the three most important known sources of NEAs are the $\nu_6$ secular resonance with Saturn, 
the 3:1 mean-motion resonance with Jupiter, 
and the Mars-crossing region. 

Once belonging to the NEAs' population the close encounters of those asteroids with the terrestrial
planets are quite frequent. 
Such characteristic makes the NEAs more accessible and so, interesting objects for exploration through space missions.

Example of a successful mission to a NEA is the Hayabusa mission, performed by the Japan Aerospace Exploration Agency and that explored the NEA (25413) Itokawa in September, 2005
\citep{b10}. The OSIRIS-REx is a mission developed by NASA designed
to return to the Earth with a sample of the NEA (101955) Bennu. The mission is planned to be launched in $2016$, and reach the NEA in 2018 \citep{b15}.
Besides those examples, we can also cite the Don Quijote program \citep{b17}, the ISHTAR program \citep{b11}, the projects SIMONE \citep{b12} 
and MARCO POLO \citep{b13}, all of them designed to send space probes to a NEA.

The multiple systems in the NEA population are especially interesting for a space mission since they increase
the observational possibilities. 
According to \cite{b16}, approximately $15\%$ of the NEAs are expected to be multiple system. 
Currently there are known 48 multiple systems in the NEAs population\footnote{Available at http://echo.jpl.nasa.gov/\texttildelow lance/binary.neas.html.  This number is constantly updated}, 
being 46 binaries, and two triple systems: 2001 SN263 \citep{b4}, and 1994 CC \citep{b14}. 

The NEA 2001 SN263 is the target of the ASTER MISSION - First Brazilian Deep Space Mission \citep{b2}. 
It is a triple system with semi-major axis $1.99$ AU, eccentricity $0.48$ and orbital inclination $6.7^{\circ}$. 
The components of the system have diameters of about 2.6 km (Alpha), 0.8 km (Beta) and 0.6 km (Gamma). With respect to the major body Alpha, 
Beta has a semi-major axis of 16.633 km - period of about 6.2 days, and Gamma has a semi-major axis of 3.804 km - period of about 0.7 days \citep{b5}. 
See Tab. 1 for more details.

The announcement of the ASTER MISSION 
has motivated a study aimed to characterize regions of stability of this system. \cite{b3} characterized the stability 
regions around the components of the triple system for the prograde case. They performed numerical integrations, for a time span of two years, of a system 
composed by the Sun, the planets Earth, Mars, Jupiter, the three components of the triple system, and thousands of particles 
randomly distributed within the internal and external regions of the triple system, with relative inclinations between 
$0^{\circ} \leqslant I \leqslant 90^{\circ}$. 

For the planar case, it were found stable regions placed very close to Alpha and Beta.
When the inclinations of the particles were considered, the effects of the Kozai mechanism became observable 
turning the whole region around the three bodies unstable for particles with relative inclination $30^{\circ}< I \leqslant 90^{\circ}$, excepted for region 1, 
where some particles survived.
It were also characterized mean motion resonances with Beta and Gamma in the internal region. It was obtained that particles with $a\approx1.8$ km, experience 
the effects of a 3:1 resonance with Gamma, while the particles with $a\approx7.8$ km and $a\approx7.9$ km, experience 
the effects of a 3:1 resonance with Beta and of a 1:3 resonance with Gamma, respectively.
For the external region of the system (besides Gamma), it was found that the region is predominantly stable, and that the variation of inclination 
does not affect the stability of the particles. 

For a space mission aimed to place the probe in the internal region of the system those results do not seem to be very interesting, since
the stable regions found are tiny and also because those are the regions more likely to have debris.
Therefore, knowing that stable regions are greater for retrograde orbits \citep{b18,b19}, here we present a complementary study, 
considering now particles orbiting the components of the system in the internal and external regions, 
with inclinations between $90^{\circ}<I\leqslant180^{\circ}$, i.e., considering the retrograde case. 

From the astronomy point of view, the analysis made before \citep{b3} for prograde orbits are more relevant because it is believed that the potential 
dust orbiting the system came from the original disk that formed the bodies, so they orbit in the prograde direction. In this case, 
retrograde orbits are not very common for natural bodies, and they are valid only for particles that are captured by the system, so in a small number.
On the other side, if the main goal is to study the system from the perspective of sending a spacecraft to visit the bodies, the 
retrograde orbits becomes an interesting option. 
The spacecraft can easily be placed in prograde or retrograde orbits, with about the same expenditure of fuel. Therefore, 
it is important to study the stability of retrograde orbits. The regions of stable retrograde orbits can be useful to 
place a spacecraft in orbits that does not need major station-keeping maneuvers in order to observe the three bodies of the system. 

Since the mission is planned to stay around one year in operation, a study of the stability of retrograde orbits are made for two years, 
that is a time that allows the designed mission with some extra time, in case the equipments last longer than planned. 
The idea is to find regions of stable retrograde orbits in locations where the prograde orbits are not stable. Those orbits allow the spacecraft 
to be in stable orbits in regions unlikely to have particles that may damage the spacecraft. Those regions are identified and shown  in detail in the present paper.
We present the regions of stability found for the retrograde case, and compare them with the regions found for the prograde case.
Our goal is to complement the results presented by \cite{b3}, given a complete scenario on the dynamic of particles in the neighborhood of the system, and thus,
provide important parameters that may guide the design of the ASTER mission. 

Only the gravitational perturbations were considered in the present paper and in the previous one.
The solar radiation pressure is a relevant mechanism to be considered in a more complete analysis since it is known that
depending on the area-to-mass ratio it may result in significant eccentricity increase, and so, affect the stability.
The analysis of how relevant this mechanism is for the specific problem of a space mission aimed to explore this triple system is the subject of a future work.
Such analysis must be guided by the results presented here and by \cite{b3}.

The structure of this paper is as follows. In Sec. \ref{sec_retrog}, we present the problem of the stability around the triple system 2001 SN263 for
the inclined retrograde case. We present the initial considerations (Sec. \ref{sec_IC}),
the adopted method  (Sec. \ref{sec_method}), and the results obtained for all the regions (Sections \ref{sec_r1} to \ref{sec_external}).
In Sec.\ref{sec_compare}, we compare the results presented here with those presented by \cite{b3}.
Based on this comparison, we also discuss the regions within the triple system that are more interesting to
place a space probe. In Sec. \ref{sec_conclusion}, we present the conclusions of the work.

\section{Inclined retrograde case}
\label{sec_retrog}
In order to achieve our goal, we first seek characterize the stable and unstable regions in the region around the components
of the triple system 2001 SN263, for the retrograde case. We based the development of this work on the initial conditions and on the method
adopted by \cite{b3}, as follows.

\subsection{Initial conditions}
\label{sec_IC}
We considered a system composed by the Sun, the planets Earth, Mars and Jupiter, the three
components of the system 2001 SN263, and by thousands of particles randomly distributed around Alpha, Beta and Gamma (internal region), and around
the whole system (external region).

The internal region was divided into four regions, defined by the approximated Hill's radius of each body (see Fig. \ref{fig_system}).
The regions 1 and 2 are the regions where Alpha is gravitationally dominant. The region 3 is the region which Beta dominates. 
The limit of this region was calculated for the problem involving Alpha and Beta, resulting in a Hill's radius of $3.4$ km. 
The region 4 is the region which Gamma theoretically dominates. 
The limit of this region was calculated for the problem involving Alpha and Gamma, resulting in a Hill's radius of $0.6$ km. 
The region 1 is limited by the orbit of Gamma, and
the region 2 is limited by the orbit of Beta. The limits of the external region are better explained
in Sec. \ref{sec_external}.

We randomly distributed particles within the regions 1, 2, 3, and in the external region. 
The region 4 was not considered since it is too small, and would become even
smaller if the perturbations from Beta were considered. 

All the particles started with an eccentricity going from $e=0.0$ until $e=0.5$, and with a random angular distribution: 
$0^{\circ} \leq f \leq 360^{\circ}$, $0^{\circ} \leq \omega \leq 360^{\circ}$ and $0^{\circ} \leq \Omega \leq 360^{\circ}$, where
$f$ is the true anomaly, $\omega$ is the argument of the pericentre and $\Omega$ is the longitude of the ascending node. 
The radial distribution of those particles depends on the region being considered. 

Regarding the inclination of the particles, we now considered the inclined retrograde case, meaning that all particles
were distributed in the internal and external regions with an inclination with respect to
the equator of Alpha given in the range $105^{\circ}\leqslant I\leqslant180^{\circ}$, taken every $15^{\circ}$.

We considered the oblateness of Alpha, with $J_2=0.013$ \citep{b5}. 
The obliquity of Alpha was also considered and was determined through the pole solution, 
presented by \cite{b5}

Beta and Gamma were assumed to be spherical bodies with radius given in Tab. 1. 
A non-sphericity of Gamma would not affect our results since we did not consider particles around this body.
A non-sphericity of Beta may change the results for region 3, especially for those particles closer to this body.
Such exception must be taken into account in further studies or in practical applications. 

\subsection{Method}
\label{sec_method}

The method consisted on numerically integrate the equations of motion of the gravitational problem of N-bodies.
The numerical integrations were performed with the Gauss-Radau numerical integrator \citep{b6}, for a time span of $2$ years. This interval corresponds to about $100$ orbital
periods of Beta, and about $1000$ orbital periods of Gamma. The time step of the numerical integrations was of $0.04$ days, for the internal and the external regions.

The physical data for Alpha, and the orbital elements of Beta and Gamma are given by \cite{b5}, for the epoch MJD 54509 in the equatorial frame of J2000 (see Tab. 1).  
The orbital elements of Alpha, Earth, Mars and Jupiter were obtained through the JPL's Horizons 
system for the same epoch, also for the equatorial frame.
The planet Saturn was not considered. Due to the short time span of the numerical integration its gravitational perturbation, 
including the effects of the $\nu_6$ secular resonance, is negligible.

\begin{figure}
\includegraphics[height=5cm]{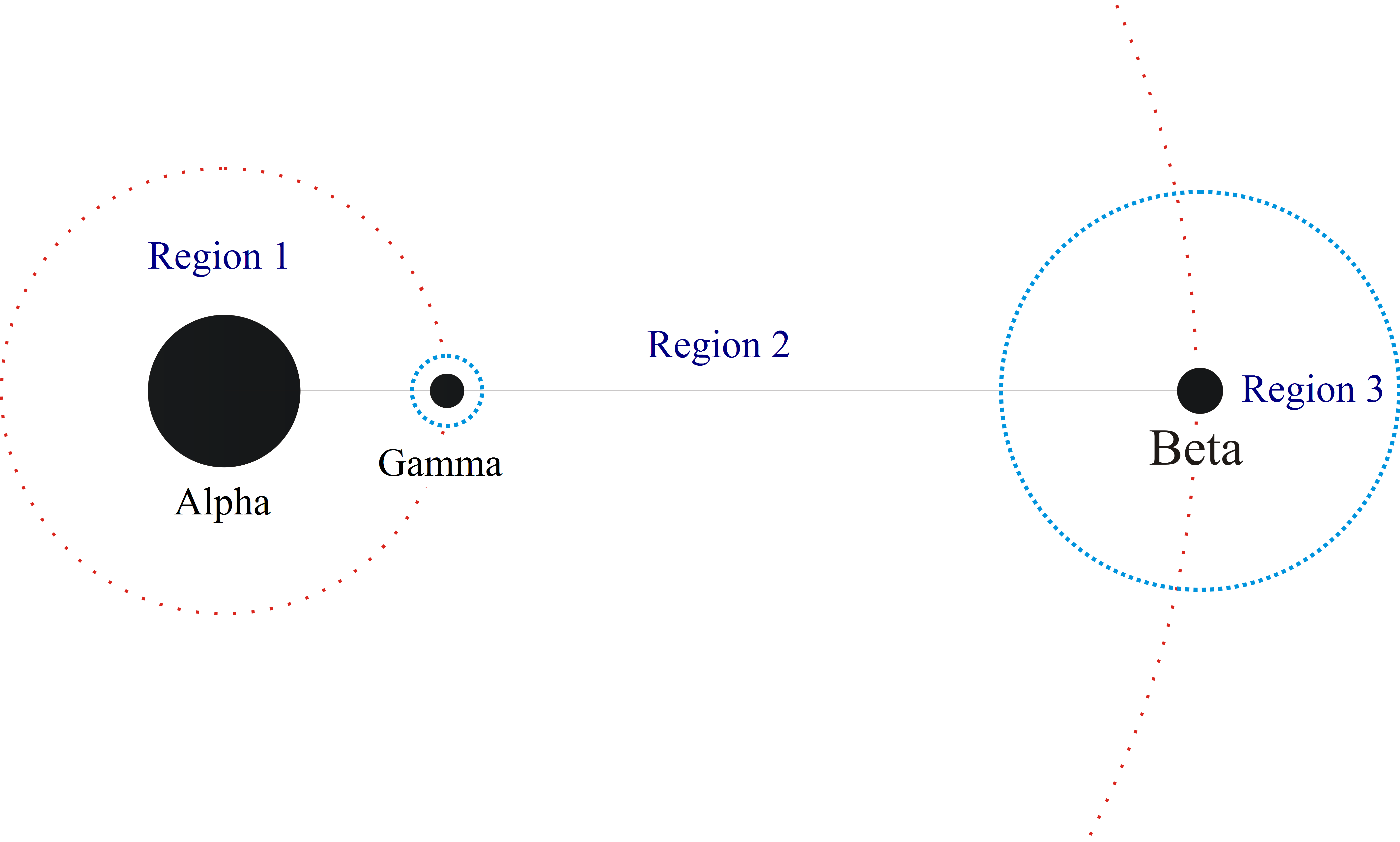}
\caption{Representation of the components of triple system 2001 SN263 and their regions of influence. The blue circles represent the Hill's radii of Beta and Gamma. 
The red dotted circles represent the collision-lines with Gamma and Beta, and by definition, the limits of the internal regions 1 and 2. Reproduced from Araujo et al. (2012).}
\label{fig_system}
\end{figure}

Throughout the numerical integrations, the particles could collide with Alpha, Beta and Gamma, or be ejected from the system. The collisions were defined by
the physical radius of the three components. The distance of ejection $d$ was defined for each region. For regions 1 and 2 it is defined by the collision-lines with Gamma and Beta,
being $d>3.804$ km for region 1 and $d>16.633$ km for region 2. The ejection limit for particles in region 3 is given by the approximated Hill's radius of Beta, being
$d>3.4$ km. Fig. \ref{fig_system} shows those limits. The ejection distance for the external region is presented in Sec. \ref{sec_external}.

The stability and instability are defined by the number of particles that survive throughout the numerical integration. 
Initial conditions that result in any loss of particles, by collision or ejection, define a stable region. Otherwise, we define the unstable regions.

Following, we discuss special features of the method described above, and we present the results obtained for the internal region (regions 1,2 and 3) and for the external region. 

\subsection{Region 1}
\label{sec_r1}

\begin{figure*}
\mbox{%
\subfigure[]{\includegraphics[height=6.2cm]{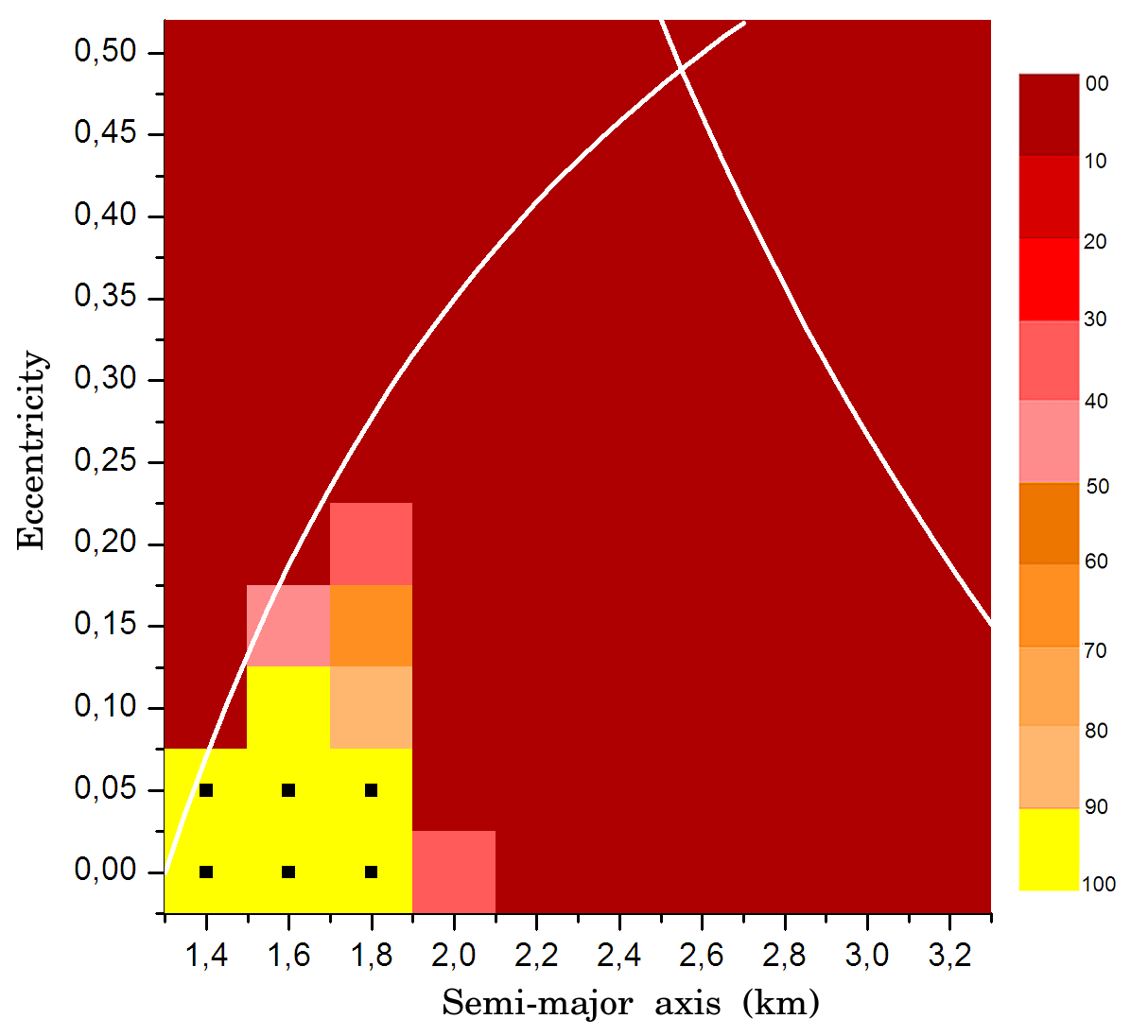}}\qquad
\subfigure[]{\includegraphics[height=6.2cm]{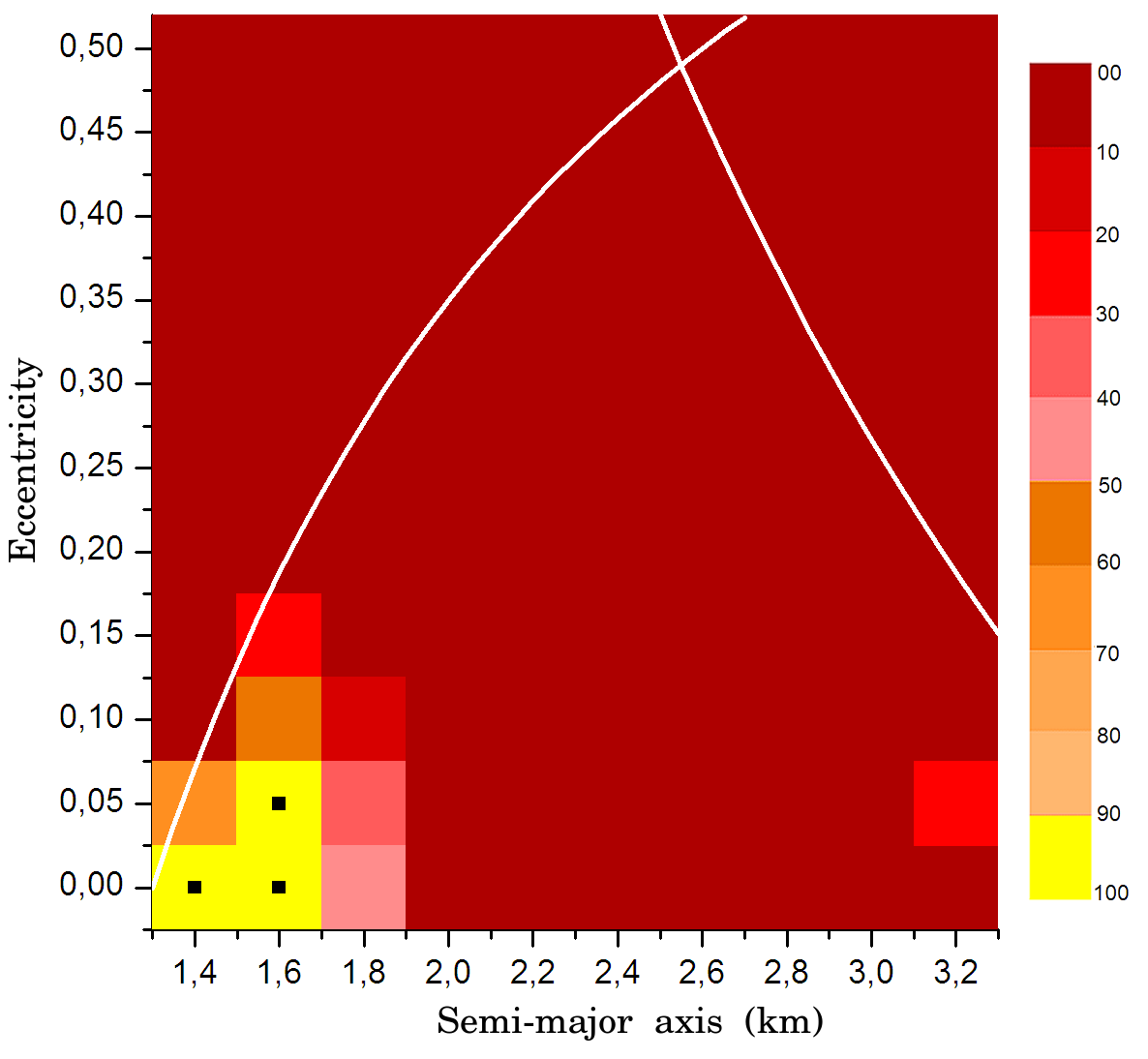}}}
\mbox{%
\subfigure[]{\includegraphics[height=6.2cm]{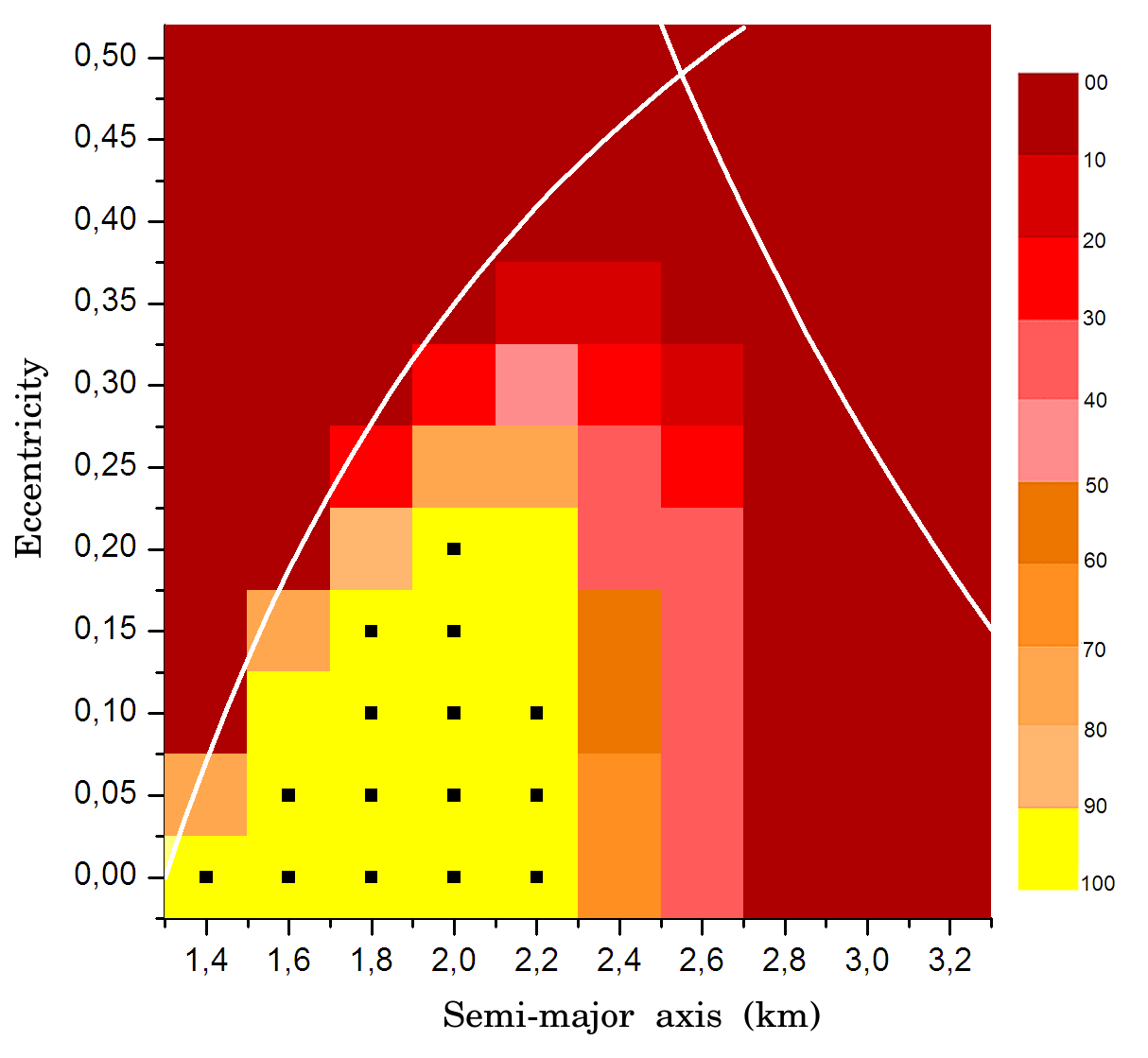}}\qquad
\subfigure[]{\includegraphics[height=6.2cm]{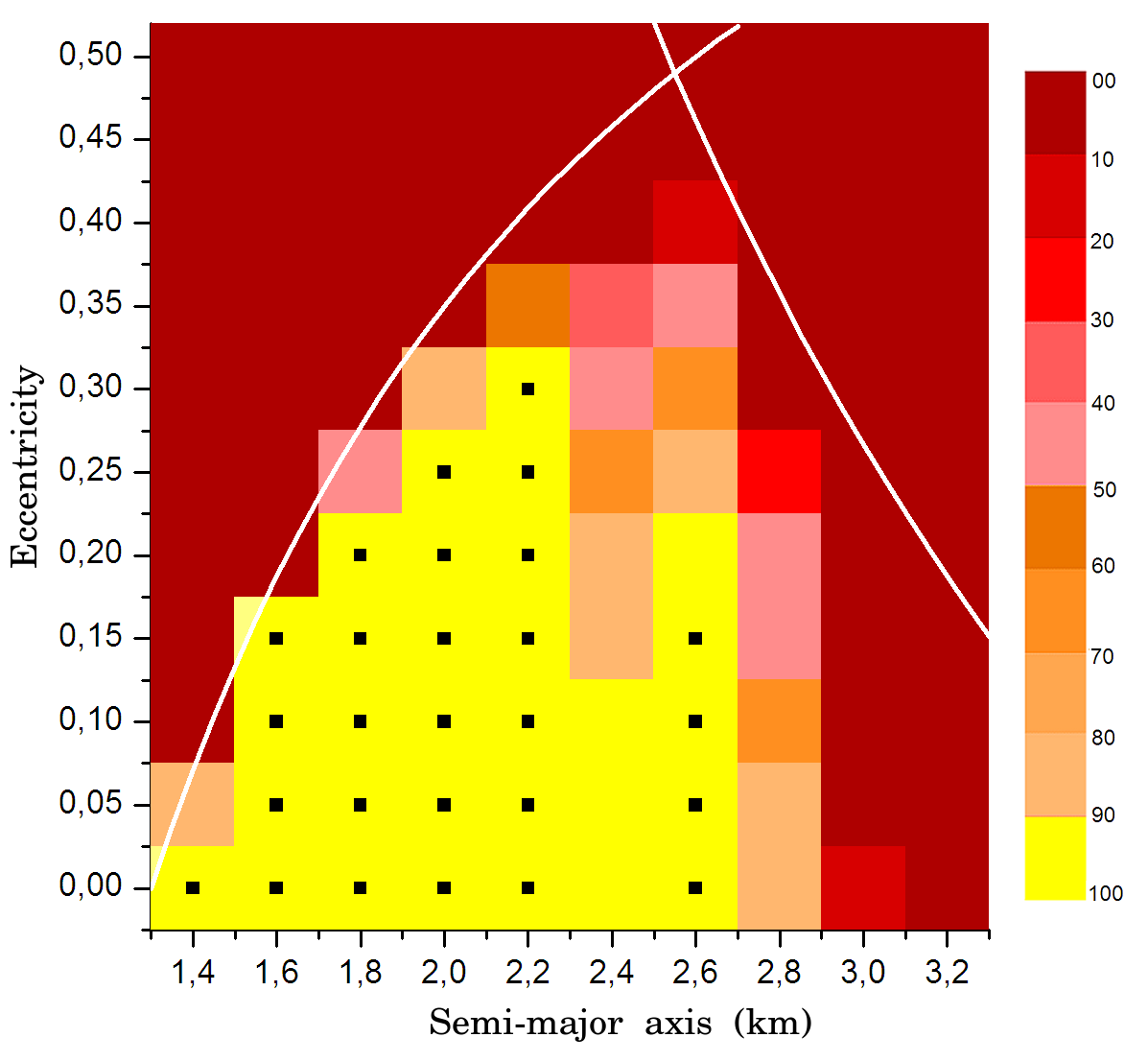}}}
\mbox{%
\subfigure[]{\includegraphics[height=6.2cm]{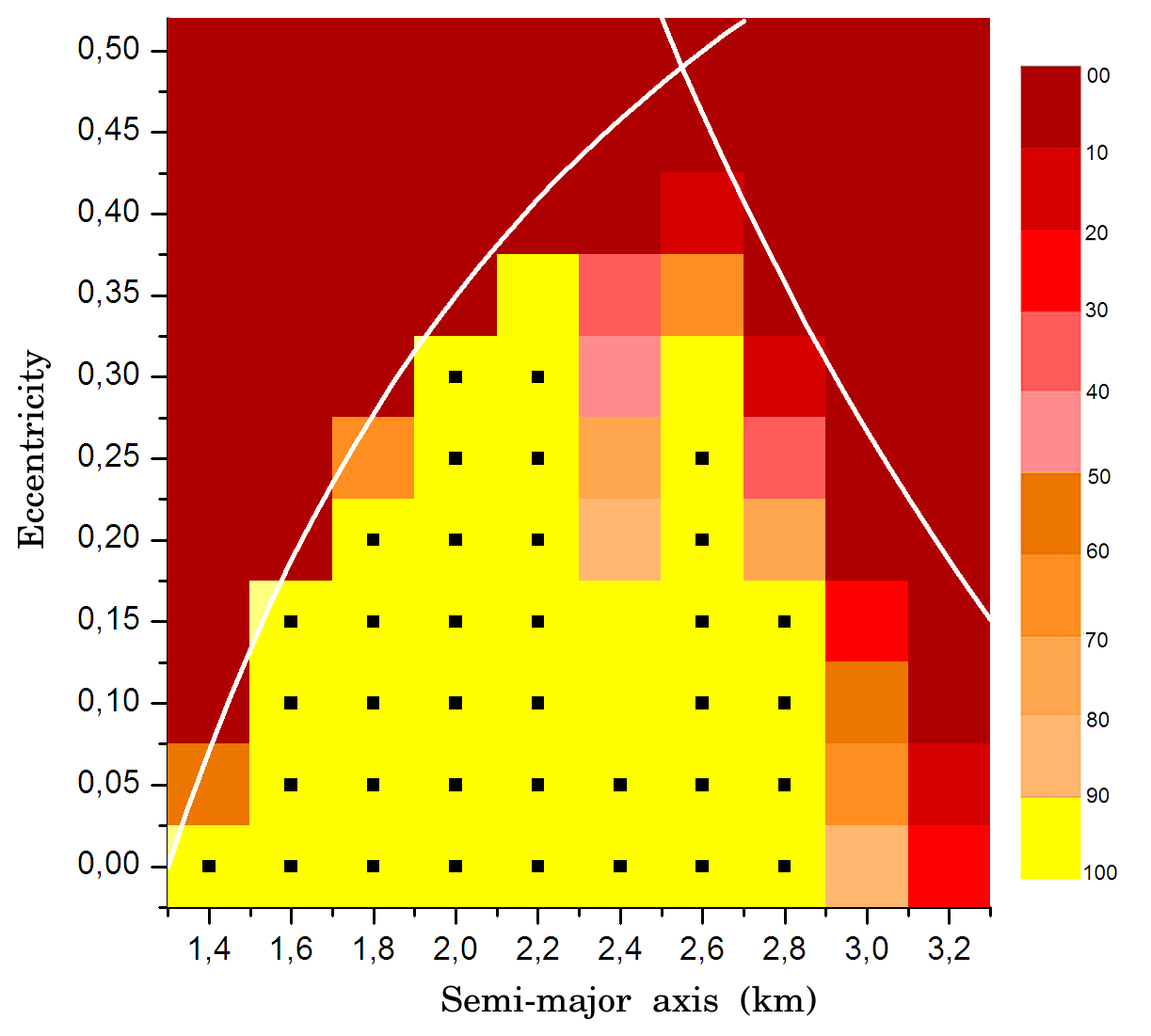}}\qquad
\subfigure[]{\includegraphics[height=6.2cm]{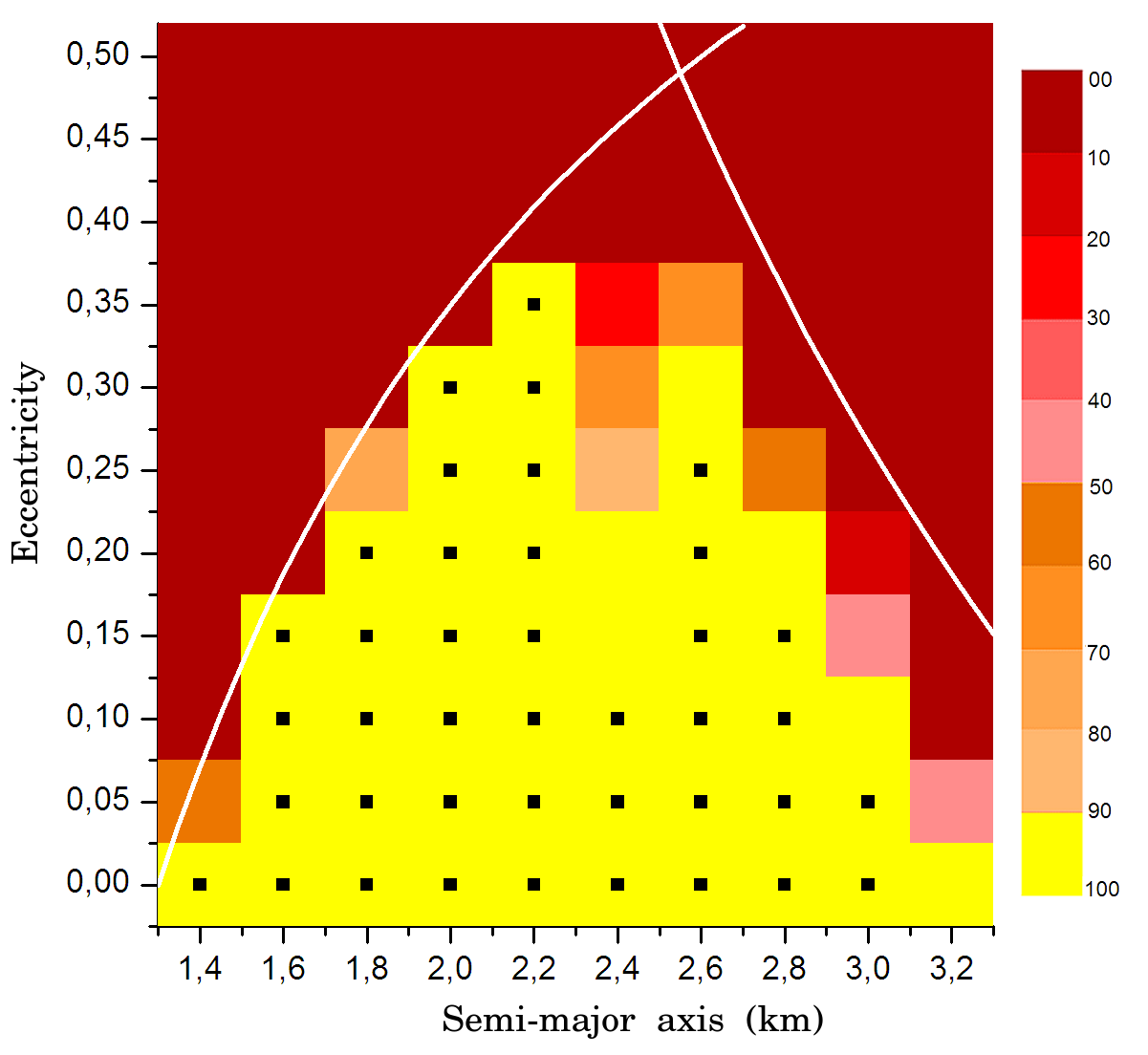}}}
\caption{Diagram of stability of region $1$ (around Alpha), for a time span of 2 years. a) $I=105^{\circ}$, b) $I=120^{\circ}$, c) $I=135^{\circ}$, 
d) $I=150^{\circ}$, e) $I=165^{\circ}$, f) $I=180^{\circ}$. The yellow boxes marked with the small
 black point indicate the cases of 100\% of survival. The white lines indicate the limits of the region. On the left is the collision-line with Alpha. 
 On the right is the collision-line with Gamma, denoting the ejection distance d.}
\label{fig_reg1}
\end{figure*}

In region 1, we considered particles orbiting Alpha with semi-major axis
$1.4 \leqslant a \leqslant 3.2$ km, taken every $0.2$ km, and with eccentricity $0.0 \leqslant e \leqslant 0.5$, taken every
$0.05$. For each combination of $a$ and $e$, we considered $100$ particles with random values of $f$, $\omega$ and $\Omega$ and with inclination
going from $105^{\circ}$ to $180^{\circ}$, taken every $15^{\circ}$. This combination of initial conditions resulted in a total of $11,000$ particles
distributed within this region, for each inclination value. The orbits were numerically integrated according to the method presented in Sec. \ref{sec_method}.

The diagrams of Fig. \ref{fig_reg1} show the stable and unstable regions found in region 1 for the retrograde case. They are composed
by a grid of semi-major axis versus eccentricity, with each combination of $a$ and $e$ represented by small squares. Each of those boxes hold the
information of 100 particles that share the same initial values of $a$ and $e$. 
The colors of the diagram indicate the percentage of survivors, going from $90\%-100\%$ of surviving particles (yellow) to 
less than 10\% of surviving particles (dark red). The small
black points indicate that $100\%$ of the particles survived (stability). The limits of the region are represented by the white lines.
They indicate the combination of $a$ and $e$ such that the pericenter of the orbits of the particles are inside the body Alpha, meaning collision (on the left),
or such that the apocenter of the orbit is beyond the ejection distance (on the right). 

In region 1, the stable region substantially increases for inclination values $I\geqslant135^{\circ}$. 
This effect is related to the Kozai mechanism \citep{b7}.
For the retrograde case \textit{the critical angle of Kozai} is $I\approx141^{\circ}$. 
Particles with inclination values lower or near this critical angle suffer the action of this mechanism, which is
known to generate oscillations of the eccentricity and of the mutual inclinations, leading to the noted instability.

The effects of the Kozai resonance is entirely noted comparing the results presented in the diagrams of Fig. \ref{fig_reg1} with the results 
presented by \cite{b3} for region 1. There is a symmetry for the prograde and retrograde cases, where particles with inclination values near the interval of the Kozai cycles
$(39,2^{\circ} \leqslant I \leqslant141^{\circ})$ presented the predicted instability. Out of this interval we have larger regions of stability. However, 
we see that for the retrograde
cases, the stable conditions are much more prominent, as expected. The instability appears only for the particles with initial conditions really close to the
limits of the regions.

It can also be noted in these diagrams the occurrence of a gap in the stable regions for particles with $a=2.4$ km. Such characteristic suggests 
the action of a resonance. This was investigated and is presented as follows.

\subsubsection{Resonance in region 1 - retrograde case}
\label{subsec_resonance}

We search for a commensurability of mean motion between particles with semi-major axis near $a=2.4$ km, with Beta or Gamma. The mean motion was calculated
through \citep{b8}:

\begin{equation}
n=\frac{G m_{p}}{a^3}\left[1+\frac{3}{2}J_{2}\left(\frac{R_{p}}{a}\right)^2\right]
\label{eq_mean}
\end{equation}
where $G$ is the gravitational constant, $m_p$ is the mass of the primary, $R_{p}$ is the radius of the primary, and $J_2$ is the gravitational harmonic
of the primary related with the oblateness of this body.

We found that particles with $a=2.398$ km have a 2:1 commensurability of mean motion with Gamma. We then search the resonant angle for  
retrograde orbits (see \cite{b20} and \cite{b21}). 

\begin{figure*}
\centering
\mbox{%
\subfigure[]{\includegraphics[height=5cm]{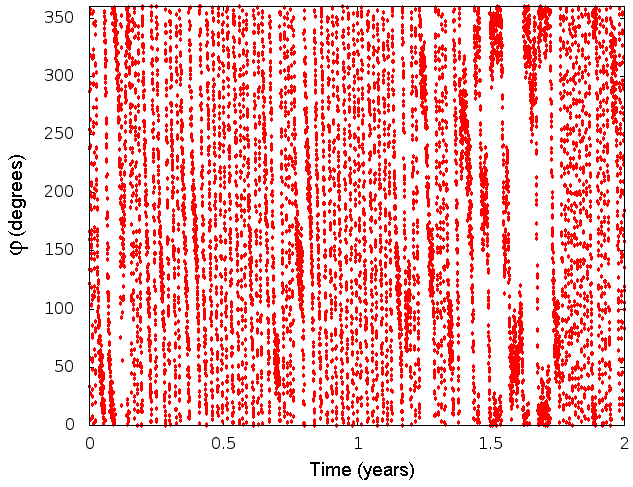}}\qquad
\subfigure[]{\includegraphics[height=5cm]{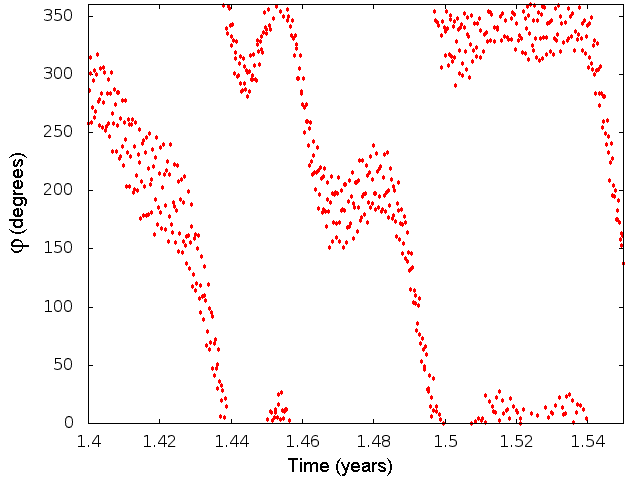}}}
\mbox{%
\subfigure[]{\includegraphics[height=5cm]{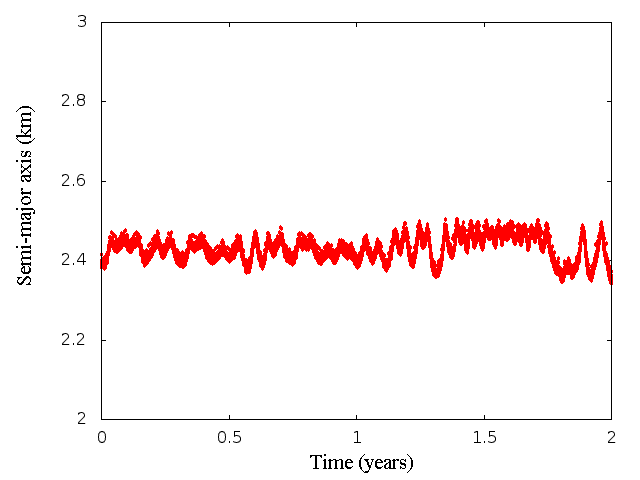}}\qquad
\subfigure[]{\includegraphics[height=5cm]{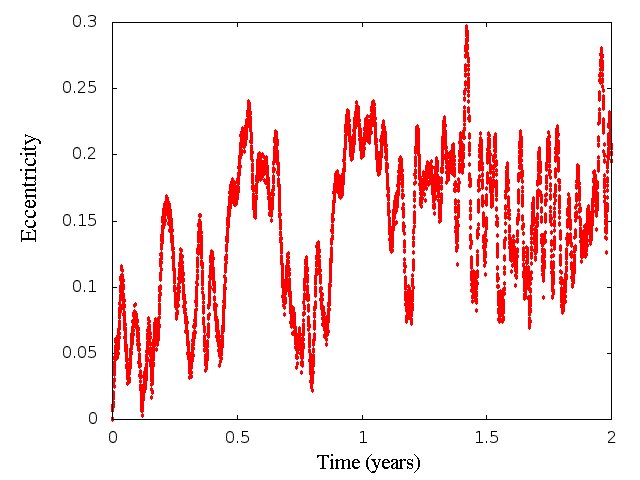}}}
\caption{Particle in region 1 with  $a=2.4$ km,  $e=0,0$ e $I=150^{\circ}$. a) Evolution of the resonant angle $\varphi=2\lambda-\lambda'-3\varpi'$  for $t=2$ years. 
b) Zoom of Figure (a) showing some of the libration regions. Time going from 1.4 to 1.55 years ($\approx79$ orbital periods of Gamma). c) Evolution of semi-major axis. d) Evolution of eccentricity.}
\label{fig_resonance}
\end{figure*}

We found that the resonant angle  $\varphi=2\lambda-\lambda'-3\varpi'$ presents an intermittent behavior (circulating
and librating), as can be seen in Figs. \ref{fig_resonance}a and \ref{fig_resonance}b.
In this equation, $\lambda'$ and $\varpi'$ is the mean longitude and the longitude of pericenter of the particle, while 
$\lambda$ is the mean longitude of Gamma.

Therefore, particles with semi-major axis near $a=2.4$ km suffer the effects of a 2:1 resonance with Gamma.
As a consequence, the particles on this region are perturbed in such way 
that the semi-major axis remains almost constant while the eccentricity increases (Figs. \ref{fig_resonance}c and \ref{fig_resonance}d), and then, the particles cross 
the collision-line of this region, giving rise to the noticed instability.

\begin{figure*}
\vspace{-0.7cm}
\subfigure{\includegraphics[height=2.5cm]{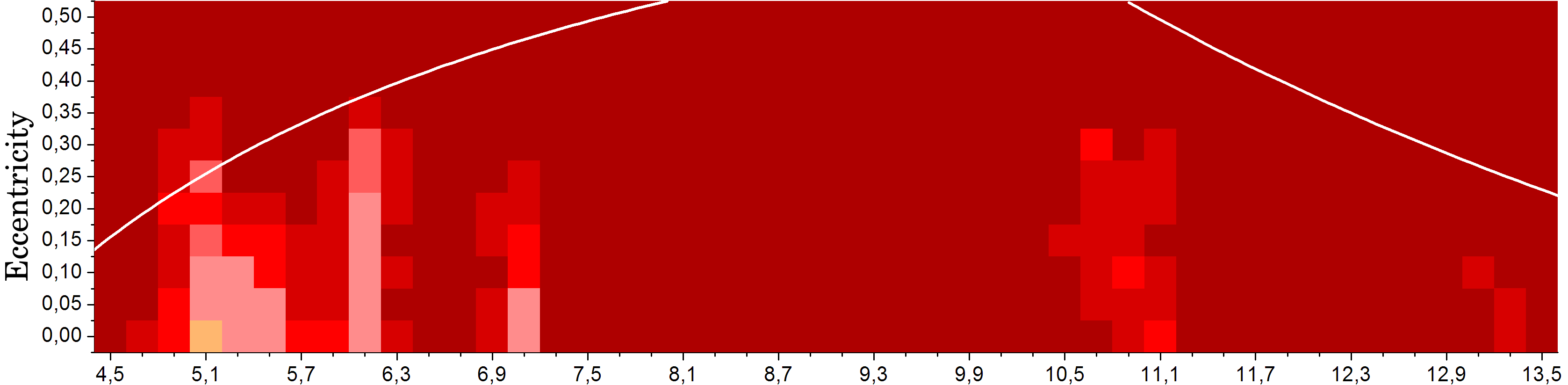}}
\subfigure{\includegraphics[height=2.5cm]{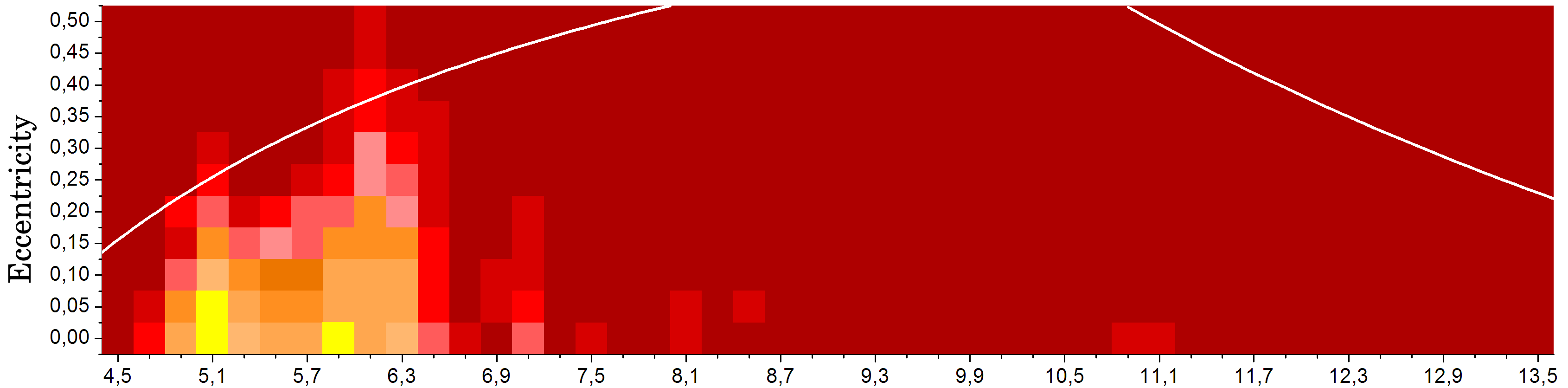}}
\subfigure{\includegraphics[height=2.5cm]{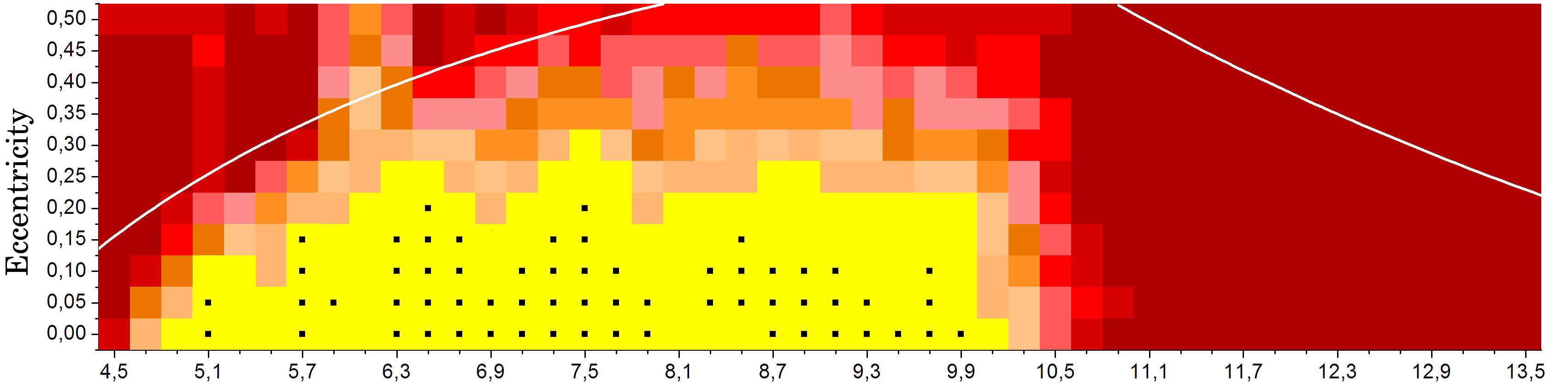}}
\subfigure{\includegraphics[height=2.5cm]{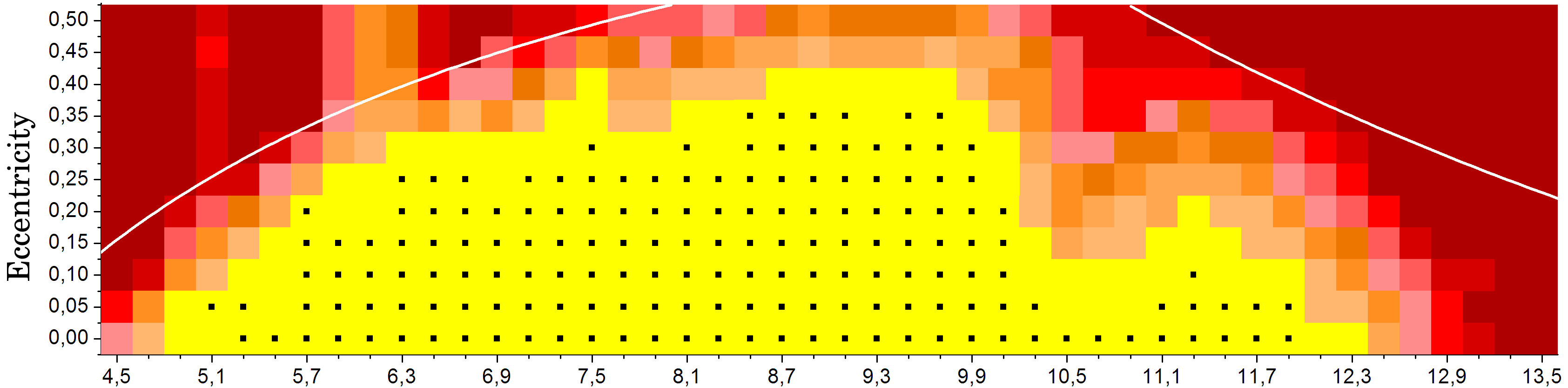}}
\subfigure{\includegraphics[height=2.5cm]{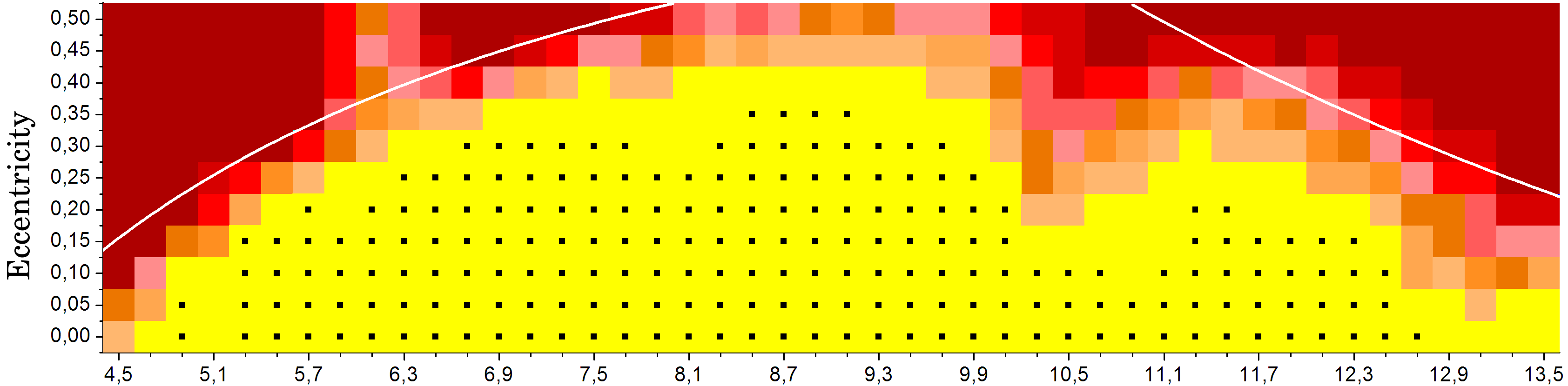}}
\subfigure{\includegraphics[height=3.55cm]{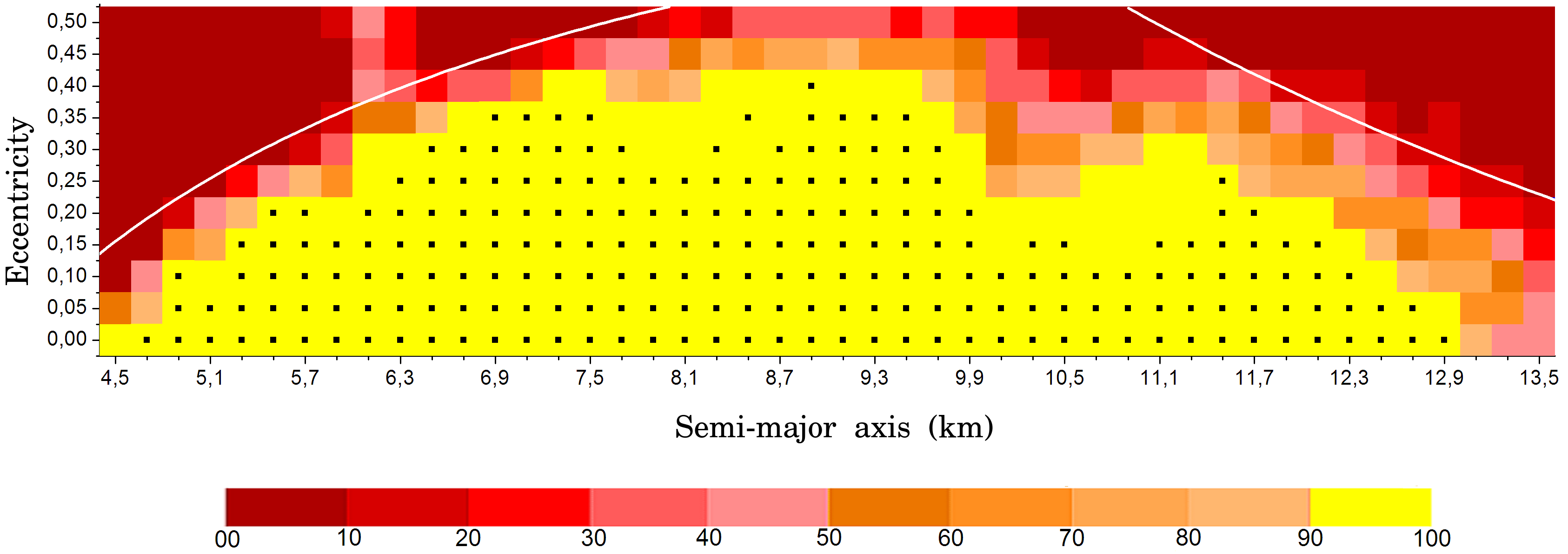}}
\caption{Diagram of stability of region $2$ (between Alpha and Beta), for a time span of 2 years for $I=105^{\circ}, I=120^{\circ},I=135^{\circ}, I=150^{\circ}, 
I=165^{\circ}$ and $I=180^{\circ}$, from top to bottom. The white lines indicate the limits of the region.The yellow boxes marked with the small
 black point indicate the cases of 100\% of survival.}
\label{fig_reg2}
\end{figure*}

\begin{figure*}
\centering
\mbox{%
\subfigure[]{\includegraphics[height=5cm]{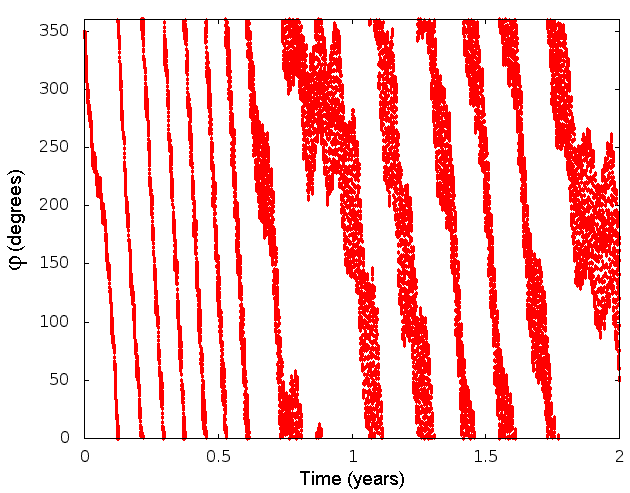}}\qquad
\subfigure[]{\includegraphics[height=5cm]{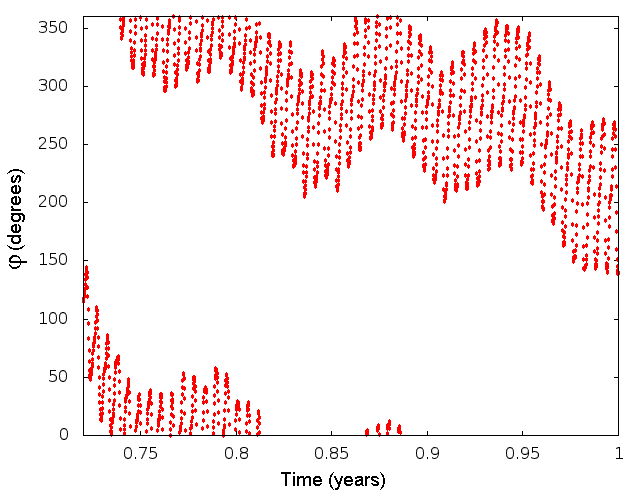}}}
\caption{Particle in region 2 with  $a=7.997$ km,  $e=0,0$ e $I=135^{\circ}$. a) Evolution of the resonant angle $\varphi=3\lambda-\lambda'-4\Omega'$  for $t=2$ years. 
b) Zoom of Figure (a) showing some of the libration regions. Time going from 0.72 to 1.0 years ($\approx 17$ orbital periods of Beta).}
\label{fig_res_r2}
\end{figure*}

\begin{figure*}
\vspace{2cm}
\mbox{%
\subfigure[]{\includegraphics[height=5.2cm]{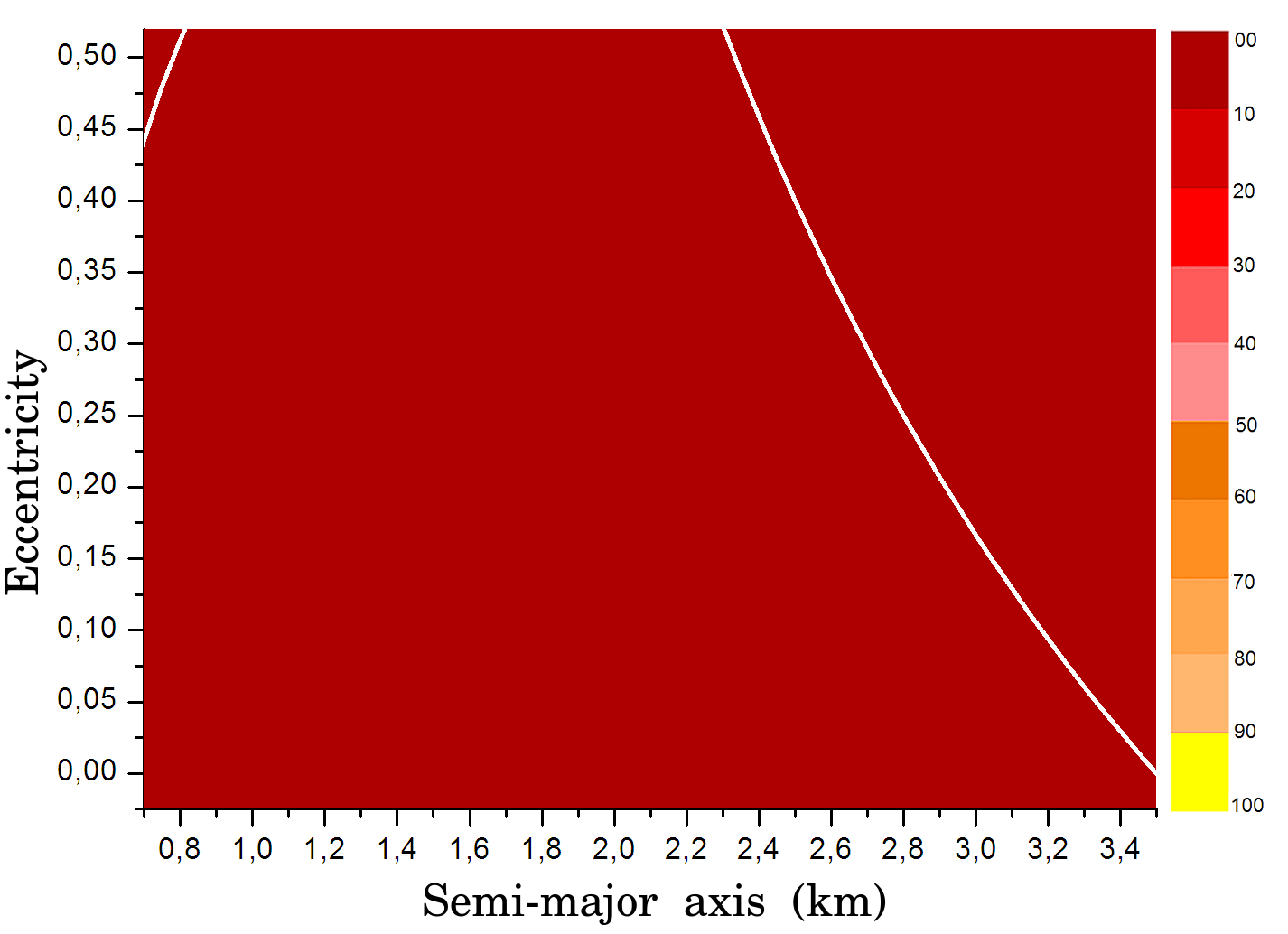}}\qquad
\subfigure[]{\includegraphics[height=5.2cm]{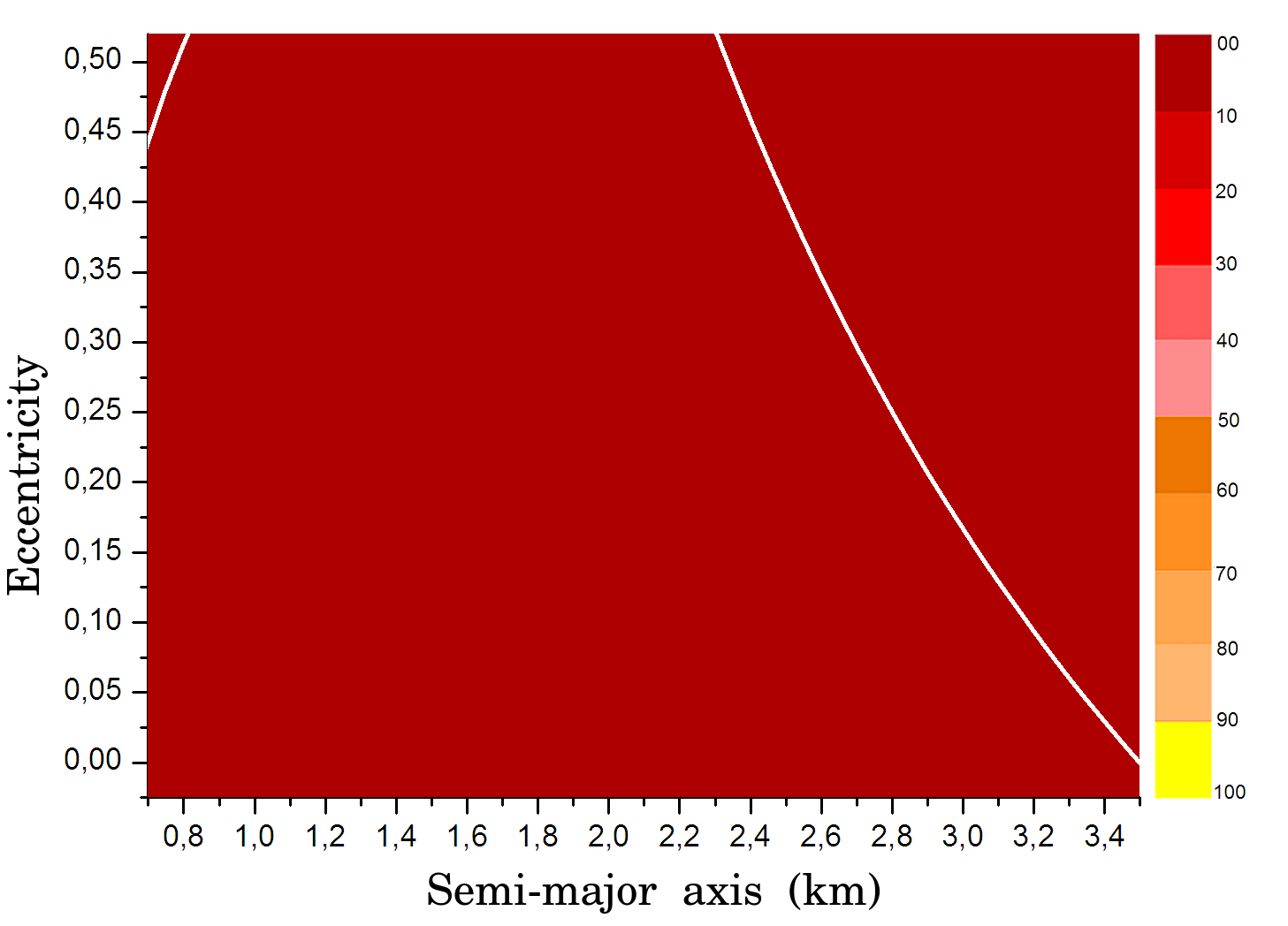}}}
\mbox{%
\subfigure[]{\includegraphics[height=5.2cm]{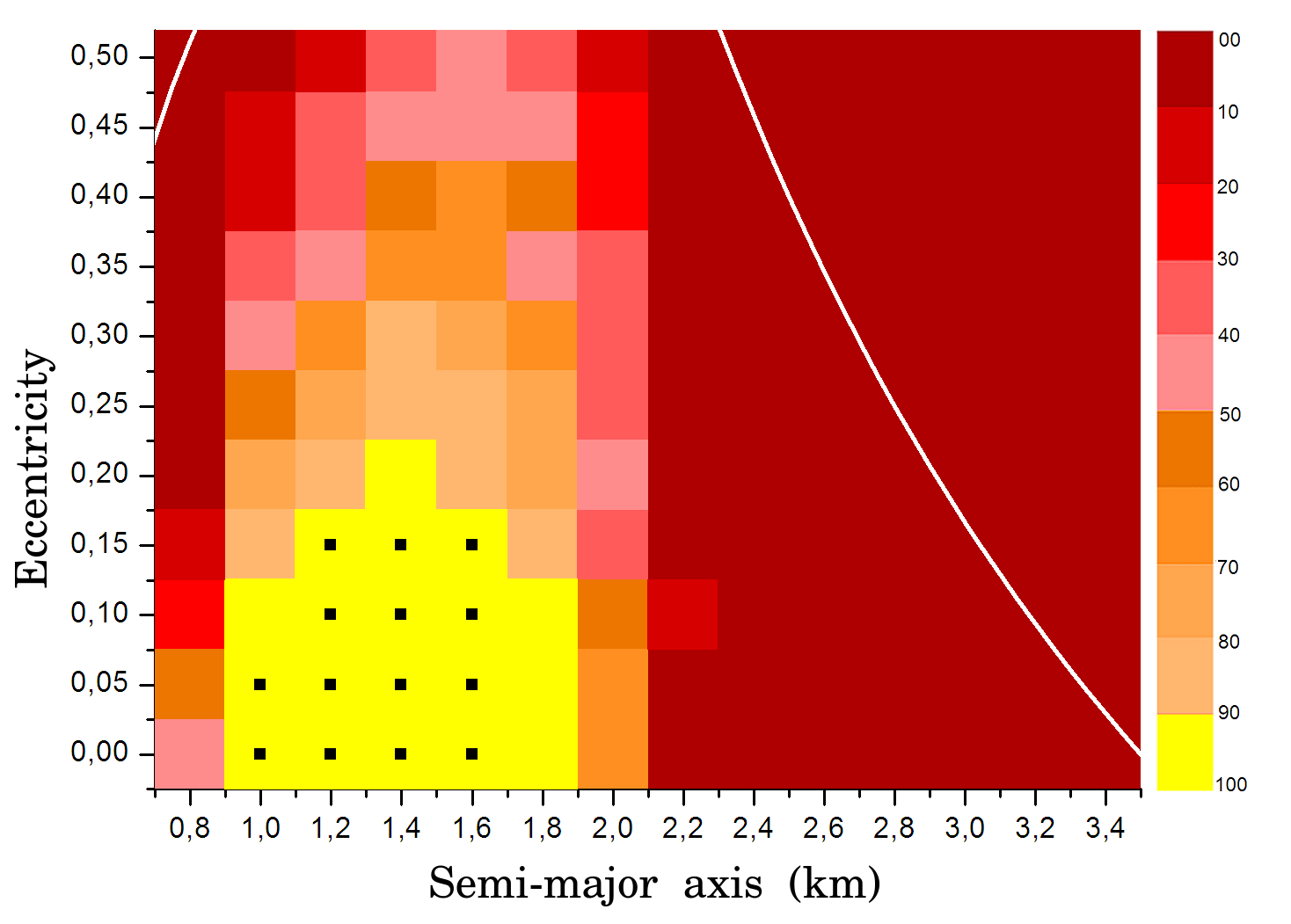}}\qquad
\subfigure[]{\includegraphics[height=5.2cm]{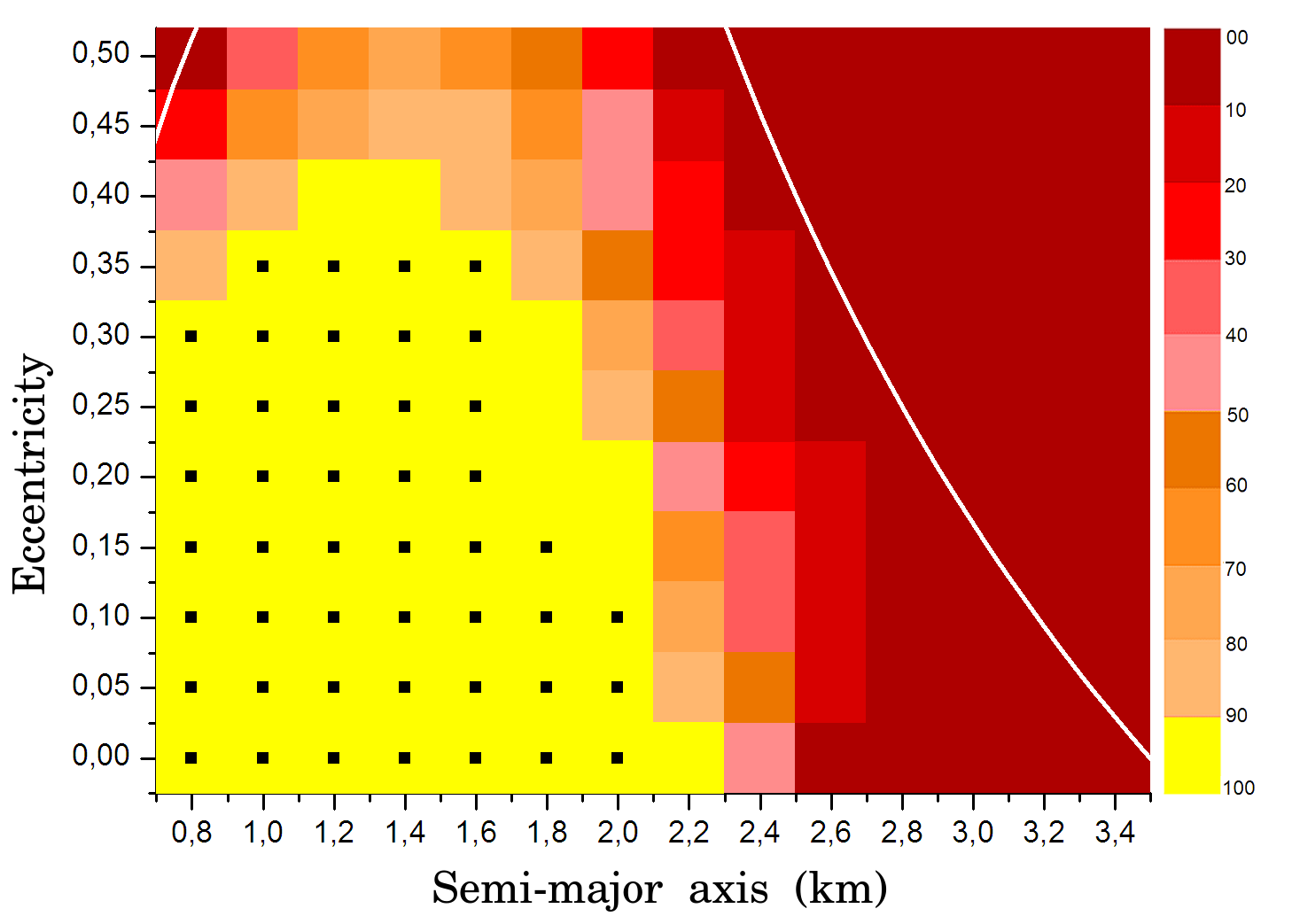}}}
\mbox{%
\subfigure[]{\includegraphics[height=5.2cm]{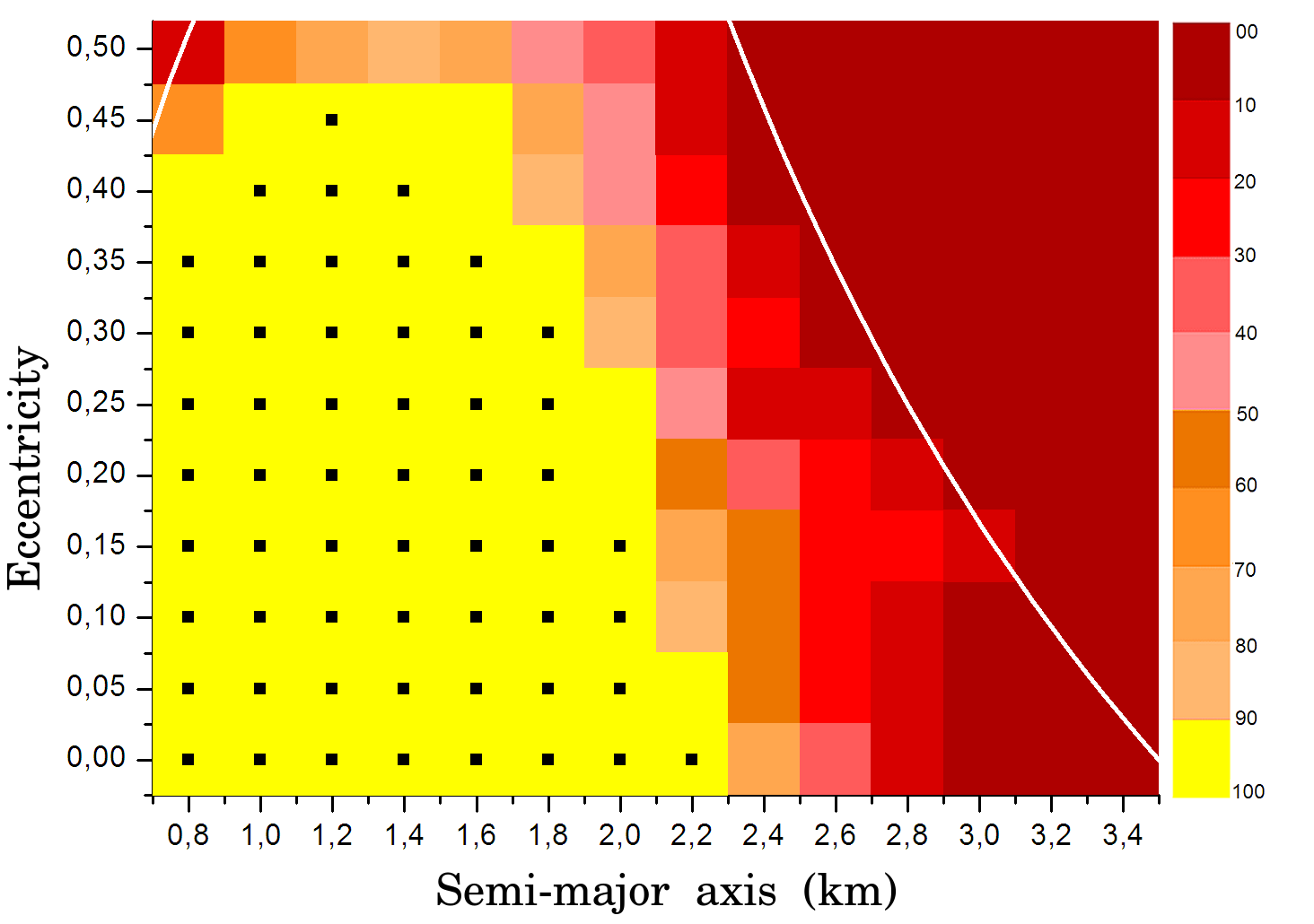}}\qquad
\subfigure[]{\includegraphics[height=5.2cm]{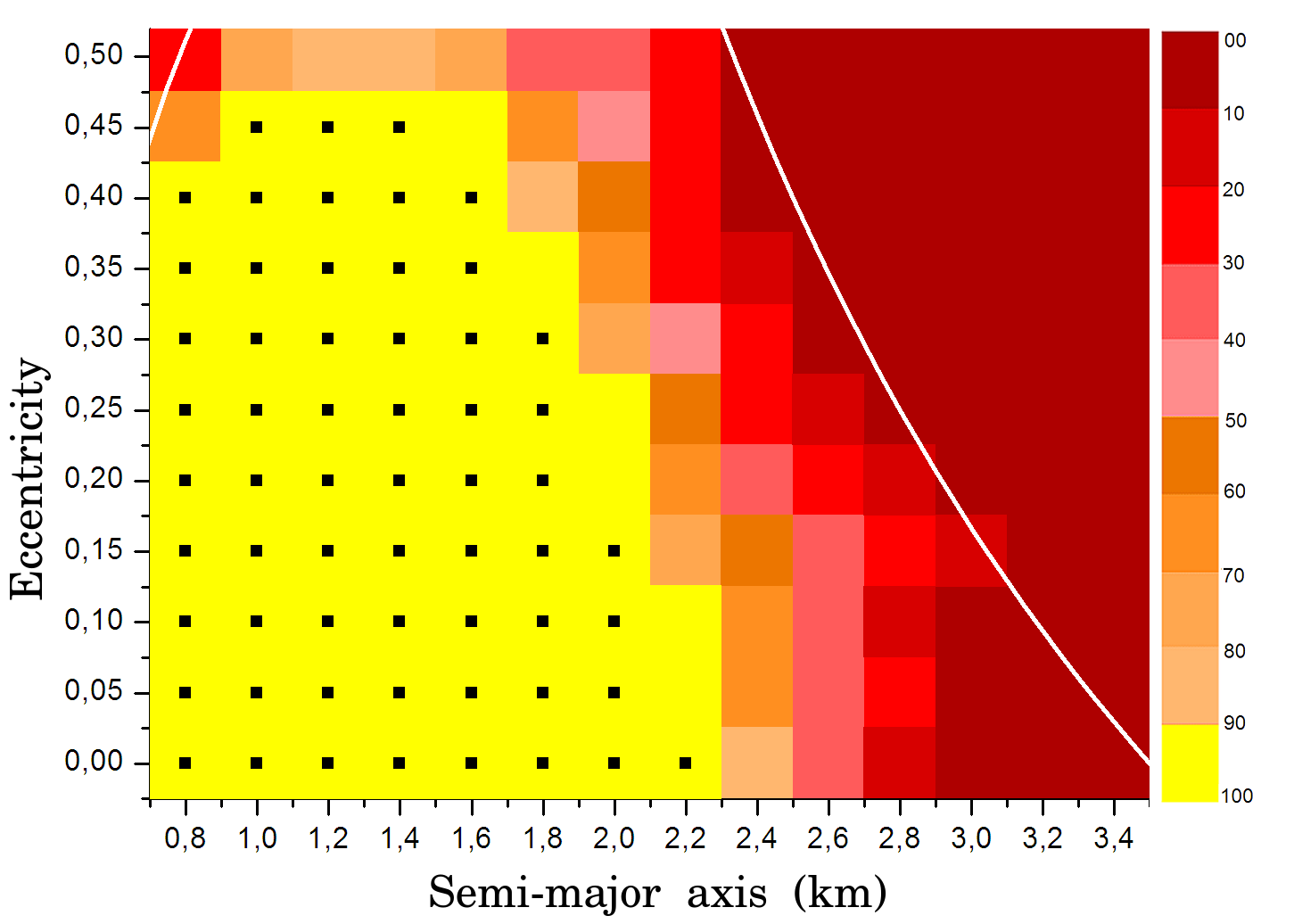}}}
\caption{Diagram of stability of region $3$ (around Beta), for a time span of 2 years. a) $I=105^{\circ}$, b) $I=120^{\circ}$, c) $I=135^{\circ}$, 
d) $I=150^{\circ}$, e) $I=165^{\circ}$, f) $I=180^{\circ}$. The white and blue lines indicate the limits of the region.The yellow boxes marked with the small
 black point indicate the cases of 100\% of survival.}
\label{fig_reg3}
\end{figure*}

\subsection{Region 2}

In region 2 the particles orbit Alpha with
$4.5 \leqslant a \leqslant 13.5$ km, taken every $0.2$ km, and with $0.0 \leqslant e \leqslant 0.5$, taken every
$0.05$. For each combination of $a$ and $e$, we considered $100$ particles with random values of $f$, $\omega$ and $\Omega$ and with inclination
going from $105^{\circ}$ to $180^{\circ}$, taken every $15^{\circ}$. This combination of initial conditions resulted in a total of $50,600$ particles
distributed within this region, for each inclination value. The orbits of these particles were integrated following the method presented in Sec. \ref{sec_method}.

The stable and unstable regions found for those conditions are presented in the diagrams of Fig. \ref{fig_reg2}. 
As presented in section \ref{sec_r1}, we adopted again a grid of semi-major axis versus eccentricity and the color code is also the same.
The limits of the region are represented by the white lines.
They indicate the combination of $a$ and $e$ such that the particles cross the collision line with Gamma, i.e., the pericenter of the orbits
are smaller than $3.804$ km (on the left), or such that the apocenter of the orbit is beyond the ejection distance $d=16.633$ km (on the right). 

We see that for the cases with $I \geq 135^{\circ}$ the region 2 presents significant stable regions. The particles with $I > 135^{\circ}$
have already crossed \textit{the critical angle of Kozai} or are near to the limit $I\approx141^{\circ}$. 
Thus, the effects of such perturbation are no longer applicable, given rise to the noted stability.
From \cite{b3}, we see that, in fact, particles with $39,2^{\circ} \leqslant I \leqslant141^{\circ}$ are under the Kozai mechanism action.

In contrast to the results presented by \cite{b3}, showing
that the region 2 is predominantly unstable for the planar and prograde cases, here, for the retrograde case, we found  
that this region is predominantly stable. For initial conditions out of the Kozai mechanism action, almost the entire region is
stable. In fact, the instability appears only for the particles with initial conditions really close to the
limits of the region (escape from the region or collisions due to crossing orbits with Gamma or Beta).

In this region we identified the action of mean motion resonances of particles with Beta and Gamma. 
Tab. $2$ brings the details of those resonances. They were found following the procedure described in Sec. \ref{subsec_resonance}.
For all of them the resonant angles librate, as exemplified in Fig. \ref{fig_res_r2}  for the resonance 3:1 with Beta.
The effects of those resonances on the particles placed in region 2 are noted through small gaps within the stable cases, 
which is more evident for higher values of eccentricities.

\begin{table}
\label{tab_resonance}
\centering
\begin{minipage}{70mm}
\centering
\caption{Mean motion resonances in region 2}
\end{minipage}
\begin{tabular}{|c c c c}
\hline
Pertuber	&Semi-major axis(km) 		& Order 	&Resonant angles$^{*}$   \\
\hline
Gamma		&$6,036$			&$1:2$     	&$\varphi=2\lambda'-\lambda-3\varpi' $      \\ 
\hline
Gamma		&$7,909$	  		&$1:3$      	&$\varphi=3\lambda'-\lambda-4\varpi $       \\
\hline
Beta		&$7,997$ 			&$3:1$      	&$\varphi=3\lambda-\lambda'-4\Omega' $ 	  \\
\hline
Beta		&$10,477$ 			&$2:1$      	&$\varphi=2\lambda-\lambda'-3\varpi' $ 	  \\
\hline
\multicolumn{4}{l}{$^{*}$ $\lambda'$ and $\varpi'$ are the mean longitude and the longitude of pericenter of the}\\
\multicolumn{4}{l}{particle. $\lambda$  and $\varpi$ are the mean longitude and the longitude of pericenter} \\
\multicolumn{4}{l}{of the satellites.}\\
\end{tabular}
\end{table}

\subsection{Region 3}

In region 3 the particles orbit Beta with
$0.8 \leqslant a \leqslant 3.4$ km, taken every $0.2$ km, and with $0.0 \leqslant e \leqslant 0.5$, taken every
$0.05$. As previously done for the other regions, we considered $100$ particles with random values of $f$, $\omega$ and $\Omega$ and with inclination
going from $105^{\circ}$ to $180^{\circ}$, taken every $15^{\circ}$, for each combination of $a$ and $e$.  
This combination of initial conditions resulted in a total of $15,400$ particles
distributed within this region, for each inclination value. The orbits of these particles were also integrated following the method presented in Sec. \ref{sec_method}.

The stable and unstable regions found for those conditions are presented in the diagrams of Fig. \ref{fig_reg3}. 
We adopted the same grid of semi-major axis versus eccentricity and color code for the diagrams. 
The limits of the region are represented by the white lines.
They indicate the combination of $a$ and $e$ such that the pericenter of the orbits of the particles are inside the body Beta meaning collision (on the left),
or such that the apocenter of the orbit is beyond the ejection distance $d=3.4$ km for region 3 (on the right).

Similar to the results for regions 1 and 2, we note that the stable regions appear only when the relative inclination is about to reach 
\textit{the critical angle of Kozai}  
$I\approx141^{\circ}$. 

The effect of the Kozai mechanism is more prominent for the particles in this region. From the diagram of Fig. \ref{fig_reg3} and
from the diagrams presented by \cite{b3}, for the region 3 and for the prograde case, we see that the stable regions completely vanish for inclination values near the interval
$39,2^{\circ} \leqslant I \leqslant141^{\circ}$.

Similar to the results obtained for regions 1 an 2, we also note a substantial increase of the stable region in region 3, when
comparing the retrograde case with the prograde case \citep{b3}. As found for regions 1 and 2, here we also found that 
the instability appears only for initial conditions really close to the limits of the region. Thus, for inclination values
out of the Kozai mechanism action, we found that retrograde orbits in region 3 are predominantly stable.

\subsection{External Region}
\label{sec_external}
We extend the analysis of the stability of the triple system 2001 SN263 to the external 
region. 

The lower limit of the external region is $d=20$ km from Alpha. This distance
corresponds approximately to the distance Alpha-Beta added to the Hill's radius of Beta.
The upper limit was determined taking into account the approximated Hill's radius of the whole system.
Considering a body with mass equal to the sum of the mass of Alpha, Beta and Gamma, and the heliocentric orbit of the system, 
\cite{b3} calculated that the Hill's radius of such body is $R_{Hill}\approx180.0$ km at the perihelion of the orbit and $R_{Hill}\approx500.0$ km at the aphelion.

According to \cite{b9} the predicted limit of stability for retrograde orbits is of about one Hill's radius. Thus, the upper limit of the external region  was chosen 
to be $d=180.0$ km (1 $R_{Hill}$ at perihelion, when the system is more perturbed). Beyond this distance the particles are considered ejected from the system.

Thus, in the external region we have particles orbiting Alpha with $20.0 \leqslant a \leqslant 180.0$ km, taken every $1.0$ km, and with $0.0 \leqslant e \leqslant 0.5$, taken every
$0.05$. As previously done, for each combination of $a$ and $e$, we considered $100$ particles with random values of $f$, $\omega$ and $\Omega$ and with inclination
going from $105^{\circ}$ to $180^{\circ}$, taken every $15^{\circ}$. This combination of initial conditions resulted in a total of $177,100$ particles
distributed within this region, for each inclination value. The orbits of these particles were integrated following the method presented in Sec. \ref{sec_method}.

\begin{figure*}
\subfigure{\includegraphics[height=1.58cm]{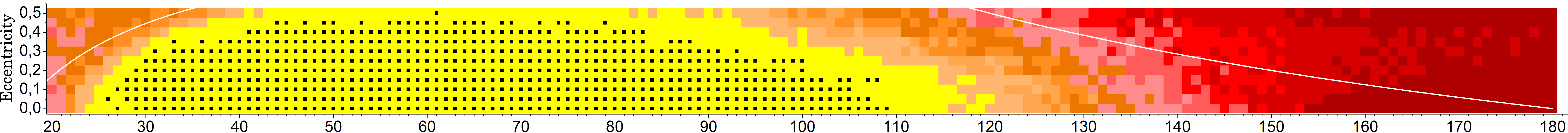}}
\subfigure{\includegraphics[height=1.64cm]{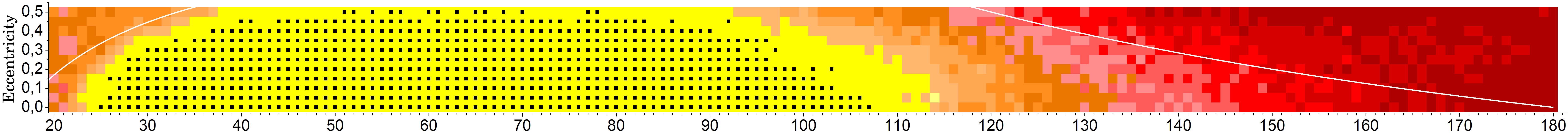}}
\subfigure{\includegraphics[height=1.62cm]{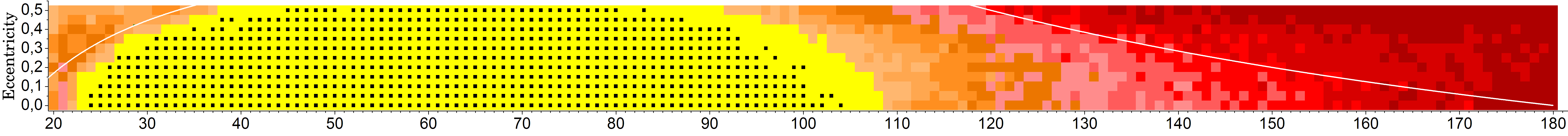}}
\subfigure{\includegraphics[height=1.64cm]{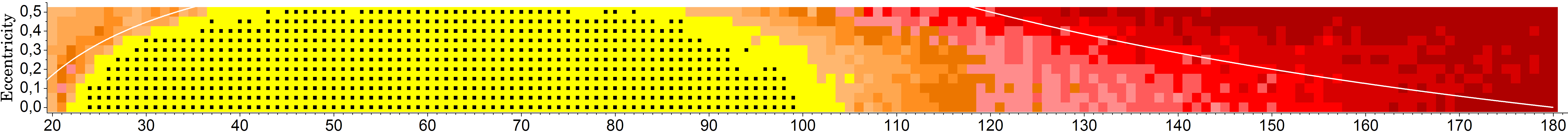}}
\subfigure{\includegraphics[height=1.64cm]{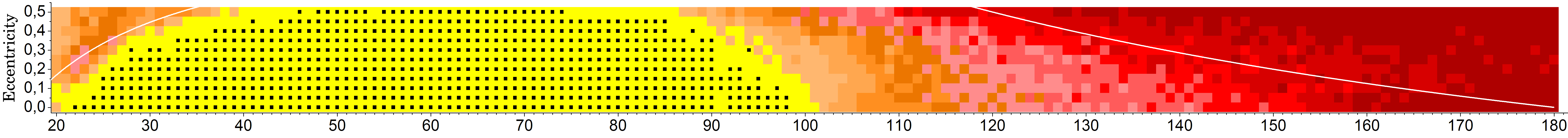}}
\subfigure{\includegraphics[height=2.64cm]{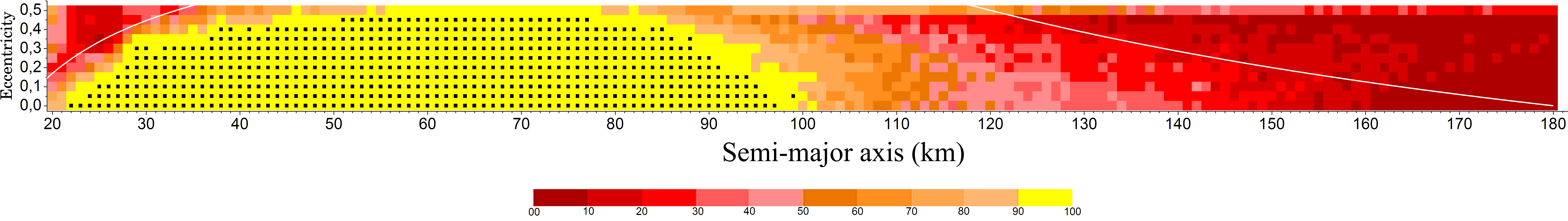}}
\caption{Diagram of stability of the external region (beyond Beta) for a time span of 2 years for $I=105^{\circ}$, $I=120^{\circ}, I=135^{\circ},I=150^{\circ}, I=165^{\circ}$, 
and $I=180^{\circ}$, from top to bottom. The yellow boxes marked with the small
 black point indicate the cases of $100\%$ of survival.}
\label{fig_regext}
\end{figure*}

The stable and unstable regions found for the external region are presented in the diagrams of Fig. \ref{fig_regext}. 
The limits of the region are represented by the white lines.
They indicate the combination of $a$ and $e$ such that the particles cross the collision line with Beta, i.e., the pericenter of the orbits
are smaller than $16.633$ km (on the left), or such that the apocenter of the orbit is beyond the ejection distance $d=180$ km (on the right). 
Comparing those diagrams with each other, we note that there is no significant difference in the stable and unstable regions. 
The instability is observable only in the border of the region.
As previously found in \cite{b3} for the planar and prograde cases, we concluded that the external region of the system is predominantly
stable for the retrograde case, and that the stability of the external region of the triple system is not affected by the inclination variation.

\section{Prograde and retrograde cases in a space mission scenario}
\label{sec_compare}

For a space mission, the most interesting regions would be those 
that are unstable for the prograde cases, but stable for the retrograde cases. 
Such configuration provide a stable region to place the mission probe for a period of 2 years with a relative retrograde orbit.
At the same time, if we take into account that possible debris located in the neighborhood of the components of the system are more 
likely to have prograde orbits,
we may have a region free of debris since they are expected to be quickly removed from the system due to the instability.

\begin{figure}
\subfigure[]{\includegraphics[height=3.0cm]{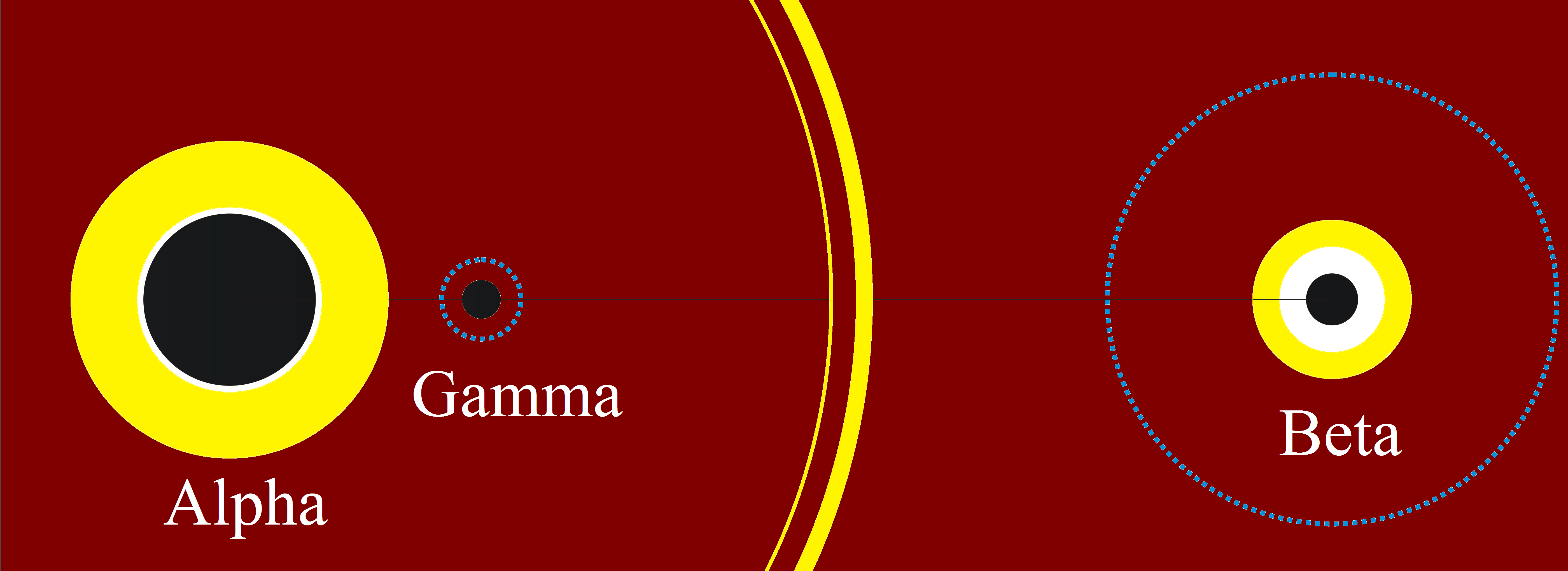}}
\subfigure[]{\includegraphics[height=3.0cm]{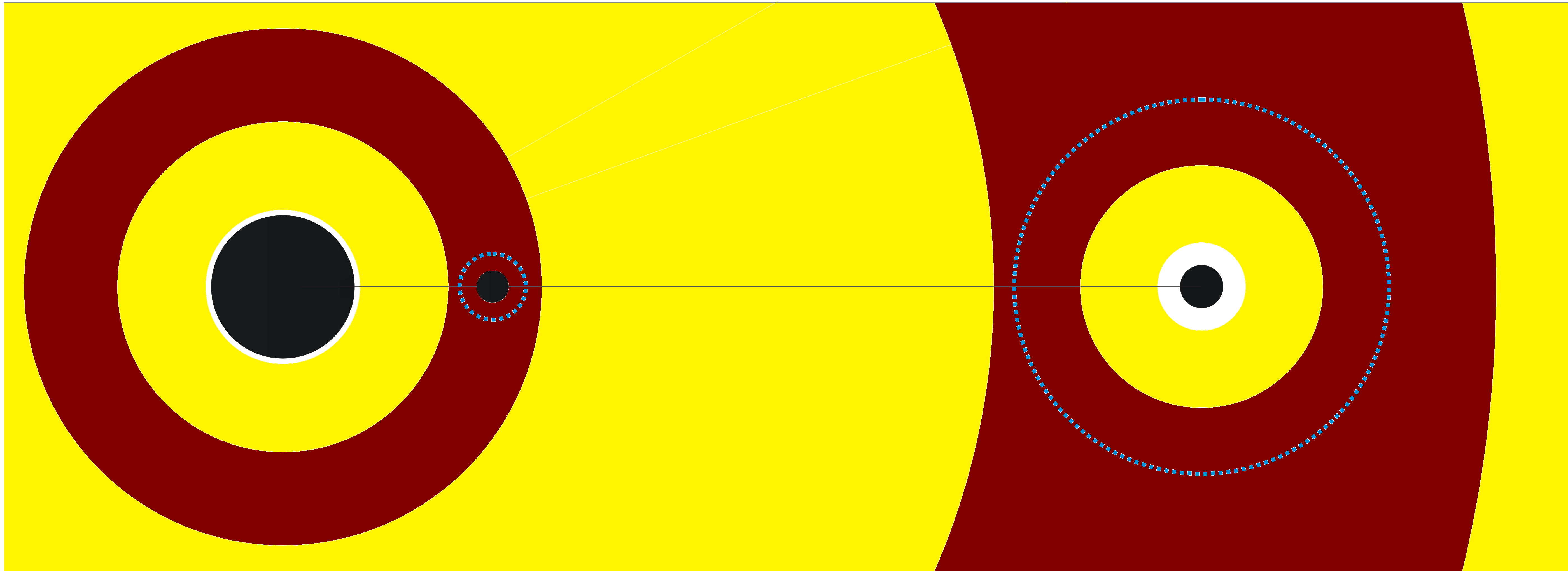}}
\subfigure[]{\includegraphics[height=3.0cm]{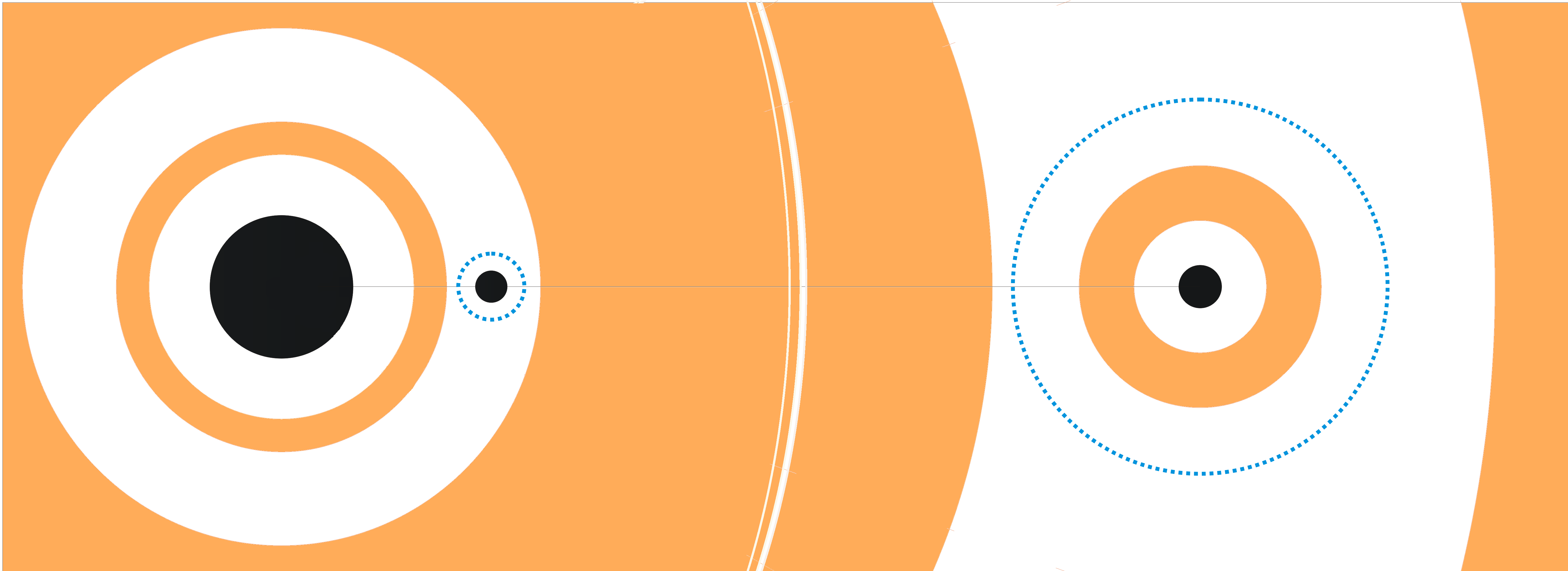}}
\caption{Comparison of the stable (yellow) and unstable (red) regions at the internal and external 
regions of the triple system 2001 SN263 for the circular case. 
a) Prograde case, reproduced from Araujo et al. (2012). b) Retrograde case. c) The difference between prograde and retrograde cases - preferred regions.}
\label{fig_compare}
\end{figure}

The diagrams in Fig. \ref{fig_compare} help us to compare the results for prograde and retrograde cases. 
They were done considering only the particles in the internal regions of the triple system
with $e=0.0$, and with $I=0^{\circ}$ (Fig.\ref{fig_compare}a), and with $I=180^{\circ}$ (Fig.\ref{fig_compare}b). These diagrams represent the radial
distribution of the particles belonging to the set of initial conditions $(a\times e)$ that after the numerical integration resulted in 
$100\%$ of survival (yellow region marked with the small black dots). 

For the planar prograde case we may consider the Hill stability, i.e., a stability criterion that consider the restricted three-body problem
and that provides a formula to calculate $\Delta$ - the mutual orbital separation -  such that the orbits remain stable \citep{b22}. 
Being Alpha the central body, we calculate $\Delta$ considering the problems: Alpha-Beta-Particle ($\mu_1=m_\beta/m_\alpha$) 
and Alpha-Gamma-Particle ($\mu_1=m_\gamma/m_\alpha$), where
the particle have the external orbit and $\Delta\approx2.4(\mu_1)^{1/3}$. 
For the Alpha-Beta-Particle problem, we found that $\Delta\approx11.8$ km, meaning that particles beyond Beta must have stable orbits
for mutual distance Beta-particle greater than this value. For the Alpha-Gamma-Particle problem, we found that $\Delta\approx2.0$ km.
Although this analytical approach does not take into account a third disturbing body and the effects of 
ressonances, we found that the results for the prograde case are in agreement with these predictions, 
as can be seen in the diagram of Fig.\ref{fig_compare}a. We see a tiny
stable zone located beyond Gamma starting at $9.1$km ($\Delta\approx5.3$ km), and beyond Beta the stable
orbits appear for particles with orbital radius of $35$ km ($\Delta\approx18.4$ km).  

From the diagram in Fig.\ref{fig_compare}b it became clear that for the retrograde case the whole region around the triple system 
2001 SN263 is essentially stable. Differently from the results obtained for the prograde case, it is clear here, for the retrograde cases, that the instability 
appears only for those particles orbiting Alpha close to the orbits of Gamma and Beta. Although it is also possible to have stable particles around Beta itself. 

In the diagram of Fig. \ref{fig_compare}c, we 
detached the preferred region with the orange color. 
We note that in regions 1 and 3 there are tiny zones around Alpha and Beta that meets the requirement of being stable for retrograde orbits and unstable for the prograde ones.
Around Alpha, we have a region with radial length of 0,6 km (region 1), and of 1,0 km around Beta (region 3). The region 2 appears as the most interesting region. 
It was characterized as unstable
for prograde orbits, but is stable when the retrograde orbits are considered. Thus, we have a region with radial length of 8.2 km that is stable for the space
probe for a period of 2 years and that must be free of debris. 

Therefore, we identify three internal regions that are very interesting to place the space probe, 
since they are stable for retrograde orbits but unstable for prograde orbits and allow the spacecraft to be in stable orbits close to the components of the system. 
The first one is around the main body, in region 1. It is very useful to place the 
spacecraft to observe the main body more closely. The second one is around the second larger body (Beta).
It is good to locate the spacecraft if the goal is to better observe this body. 
The third one is in the regions between the two satellite bodies (region 2). It is a very interesting region to keep the spacecraft to observe the three bodies of the system. 
An elliptical orbit with periapsis near the lower limit of this region and apoapsis near the upper limit of this region would 
allow the spacecraft to pass closer to both secondary bodies in a single stable orbit that stays inside a region that is 
free of concentration of particles.

The external region are essentially stable for both, prograde and retrograde orbits. Nevertheless, we can detach a region of about $7$ km,
going from $a=27$ km to $a=34$ km, beyond Beta, that meet the \textit{preferred region} requirement. This would be a very interesting region
to place a space probe. There we may have a clean area with stable retrograde orbits around Alpha that allow a close observation of the three
components of the system.

\section{Conclusions}
\label{sec_conclusion}

In the present paper we analyze stability regions of the triple system 2001 SN263, for retrograde orbits in the internal and external regions of the system. 
Based on those results, we detached the preferred regions, i.e, internal and external regions that are interesting to place a space probe aimed to explore
this system.

We applied the method adopted by \cite{b3}, but now considering  particles with retrograde orbits, i.e., particles with $90^{\circ}<I\leqslant180^{\circ}$ relative
to the equator of Alpha, within and around the entire system. We analyzed the behavior of such particles to determine the stable regions
for the retrograde case, and we compared our results with the planar and prograde cases

Comparing our results with the planar and prograde cases we noted, as expected \citep{b18,b19}, a substantial increase of the stable region for both, internal and external regions
when the retrograde orbits were considered. 

We found that nearly the whole internal region is stable, especially for low eccentricities values. 
The instability appears only in the border of the regions, or when
the inclination of the particles are such that they are under the Kozai mechanism action. 
The most prominent result was found for region 2. Comparing the simulations for particles with $I=0^{\circ}$ with those with $I=180^{\circ}$, we see that
the region that is essentially unstable for the prograde case, became practically fully stable for the retrograde case. 

We found that the external region is also predominantly stable. The instability appears only in the borders of the region. 
The variation of inclination does not affect the stability of the particles.

Our results showed that the retrograde orbits are an interesting option for the ASTER mission. We 
detached at least four regions (three internal and one external) that can be very useful to this mission. 
All of them allow the space probe to closely observe the components of the system in stable retrograde orbits
situated in regions expected to be free of debris. 
Those results provide important information about the stability around the components of the system and substantially contribute to the design of the ASTER mission. 

\section{Acknowledgments}
This work was funded by INCT - Estudos do Espa\c co, CNPq and FAPESP (proc. 2011-08171-3).  This support is
gratefully acknowledged. We also thank Ernesto Vieira Neto for his contribution with the numerical integrator.

\renewcommand{\refname}{REFERENCES}

\label{lastpage}


\begin{thebibliography}{1}
\bibitem[\protect\citeauthoryear{}{}]{b22}


\bibitem[\protect\citeauthoryear{Amata,}{2009}]{b13}Amata, G.B.; Marco Polo Mission - Executive Summary. 15p. 2009.

\bibitem[\protect\citeauthoryear{Araujo et al.,}{2012}]{b3}Araujo, R. A. N.; Winter, O. C.; Prado, A. F. B. A.; Sukhanov, A., Stability regions around the components of the triple system 2001 SN263. MNRAS, Vol. 423, Issue 4, pp. 3058-3073..

\bibitem[\protect\citeauthoryear{Bottke \& Melosh}{1996a}]{b16}Bottke, W. F., Melosh, H. J. Binary Asteroids and the Formation of Doublet Craters. Icarus, v. 124, pp. 372-391, 1996a.

\bibitem[\protect\citeauthoryear{Brozovic et al.,}{2009}]{b14} Brozovic, M.; Benner, L. A. M.; Nolan, M. C, et al. (136617) 1994 CC. IAU Circ., 9053, 2, 2009.

\bibitem[\protect\citeauthoryear{D'arrigo,}{2003}]{b11}D'arrigo, P. The ISHTAR Mission Executive Summary for Publication
on ESA Web Pages. Madrid. 8p. Technical report. 2003.

\bibitem[\protect\citeauthoryear{Domingos et al.,}{2006}]{b9} Domingos, R. C.; Winter, O. C.; Yokoyama, T. Stable satellites around extrasolar giant planets. MNRAS, V. 373, p.  1227-1234, 2006. 
 
\bibitem[\protect\citeauthoryear{Everhart,}{1985}]{b6}Everhart, E. An efficient integrator that uses Gauss-Radau spacings. In Dynamics of comets: Their origin and evolution, Eds. A. Carusi Carusi and G. B. Valsecchi, D.Reidel Publishing Company (Holanda), p. 185-202, 1985.

\bibitem[\protect\citeauthoryear{Fang et al.,}{2011}]{b5}Fang, J. et al. Orbits of near-earth asteroid triples 2001 SN263 and 1994 CC: properties, origin, and evolution. Astronomical Journal. Volume 141, Issue 5, 2011. 

\bibitem[\protect\citeauthoryear{Galvez,}{2003}]{b17} Galvez,  A. et al. Near Earth Objects Space Mission Preparation: Don Quijote Mission Executive Summary. Madrid. 9p. Technical report, 2003.

\bibitem[\protect\citeauthoryear{Gladman,}{1993}]{b22}Gladman, B. Dynamics of systems of two close planets. Icarus, vol. 106, p. 247, 1993

\bibitem[\protect\citeauthoryear{Hamilton \& Burns}{, 1991}]{b18} Hamilton, D. P.; Burns, J. A. Orbital stability zones about asteroids. Icarus, V. 92, p. 118-131, 1991.

\bibitem[\protect\citeauthoryear{Hamilton \& Krivov}{, 1997}]{b19} Hamilton, D. P.; Krivov, A .V. Dynamics of Distant Moons of Asteroids. Icarus, V. 128, Issue 1, p. 241-249, 1997.

\bibitem[\protect\citeauthoryear{Hergenrother et al.,}{2014}]{b15}Hergenrother, Carl W. et al., The Design Reference Asteroid for the OSIRIS-REx Mission Target (101955) Bennu. eprint arXiv:1409.4704, 116 pages, 2014.

\bibitem[\protect\citeauthoryear{Kozai,}{1962}]{b7} Kozai, Y. Secular perturbation od asteroids with high inclination and eccentricity, The astronomical journal, v. 67, n.9, 
p. 591-598, 1962.

\bibitem[\protect\citeauthoryear{Morais \&  Giuppone}{2012}]{b20} Morais, M. H. M.; Giuppone, C. A. Stability of prograde and retrograde planets in circular binary systems. MNRAS, V. 424, Issue 1, p. 52-64, 2012.

\bibitem[\protect\citeauthoryear{Morais \& Namouni}{2013}]{b21}Morais, M. H. M.; Namouni, F. Retrograde resonance in the planar three-body problem. CMDA, V. 117, Issue 4, p.405-421, 2013.

\bibitem[\protect\citeauthoryear{Morbidelli et al.,}{2002}]{b1} Morbidelli, A.; Bottke, W. F., Jr.; Froeschl\'e, Ch.; Michel, P. Origin and Evolution of Near-Earth Objects. Asteroids III, W. F. Bottke Jr., A. Cellino, 
P. Paolicchi, and R. P. Binzel (eds), University of Arizona Press, Tucson, p.409-422 ,2002.

\bibitem[\protect\citeauthoryear{Murray and Dermott,}{1999}]{b8} Murray, D.C.; Dermott, S.F. Solar System Dynamics. Cambridge University Press. 1999. 

\bibitem[\protect\citeauthoryear{Nolan et al.,}{2008}]{b4} Nolan, M.C. et al., Arecibo radar imaging of 2001 SN263: a near-earth triple asteroid system. Asteroids, Comets, Meteors, n. 8258, 2008.

\bibitem[\protect\citeauthoryear{Sukhanov et al.,}{2010}]{b2}Sukhanov, A.A., Velho, H.F.C., Macau, E.E., Winter, O.C. The Aster Project: Flight to a Near Earth Asteroid. Cosmic Research, 2010, Vol. 48, No. 5 pp. 443-450.

\bibitem[\protect\citeauthoryear{Wells,}{2003}]{b12} Wells, N. SIMONE NEO Mission Study Executive Summary. Hampshire.11p. Technical report. 2003.

\bibitem[\protect\citeauthoryear{Yoshikawa,}{2006}]{b10} Yoshikawa,  M. et al., Technologies for future asteroid exploration: What we learned from hayabusa mission. Spacecraft Reconnaissance of Asteroid and Comet Interiors, n.3038,2006.

\end{thebibliography}
\end{document}